\documentclass[a4paper,11pt]{article}
\pdfoutput=1

\usepackage{jheppub_kim}
\usepackage[T1]{fontenc}
\usepackage{amssymb}
\usepackage{graphicx}
\usepackage{amsmath}
\usepackage{tensor}
\usepackage{hyperref}
\usepackage{epstopdf}
\usepackage{extarrows}
\usepackage{xcolor}

\newcommand{\td}{\text{d}}

\def\I {\hat{\mathbb{I}}}
\def\x {\hat{x}}
\def\O {\hat{O}}

\def\Tr {{\text{Tr}}}

\def\C {\mathcal{C}}
\def\F {\tilde{F}}
\def\V {\hat{V}}
\def\W {\hat{W}}
\def\U {\hat{U}}

\newcommand{\kyr}[1]{{\color{black}{#1}}}

\begin{document}

\title{Principles and symmetries of complexity in quantum field theory}

\author[a]{Run-Qiu Yang,}
\author[b,c]{Yu-Sen An,}
\author[d,e]{Chao Niu,}
\author[f]{Cheng-Yong Zhang,}
\author[d]{Keun-Young Kim}

\emailAdd{aqiu@kias.re.kr}
\emailAdd{anyusen@itp.ac.cn}
\emailAdd{chaoniu09@gmail.com}
\emailAdd{zhangchengyong@fudan.edu.cn}
\emailAdd{fortoe@gist.ac.kr}

\affiliation[a]{Quantum Universe Center, Korea Institute for Advanced Study, Seoul 130-722, Korea}
\affiliation[b]{Institute of Theoretical Physics, Chinese Academy of Science, Beijing 100190, China}
\affiliation[c]{School of physical Science, University of Chinese Academy of Science, Beijing 100049, China}
\affiliation[d]{ School of Physics and Chemistry, Gwangju Institute of Science and Technology,
Gwangju 61005, Korea
}
\affiliation[e]{Department of Physics and Siyuan Laboratory, Jinan University, Guangzhou 510632, China}
\affiliation[f]{Department of Physics and Center for Field Theory and Particle Physics, Fudan University, Shanghai 200433, China}

\abstract{
Based on general and minimal properties of the {\it discrete} circuit complexity, we define the complexity in {\it continuous} systems in a geometrical way.  We first show that the Finsler metric naturally emerges in the geometry of the complexity in continuous systems. Due to fundamental symmetries of quantum field theories, the Finsler metric is more constrained and consequently, the complexity of SU($n$) operators is uniquely determined as a length of a geodesic in the Finsler geometry. Our Finsler metric is bi-invariant contrary to the right-invariance of discrete qubit systems. We clarify why the bi-invariance is relevant in quantum field theoretic systems. After comparing our results with discrete qubit systems we show most results in $k$-local right-invariant metric can also appear in our framework.  Based on the bi-invariance of our formalism, we  propose a new interpretation for the Schr\"{o}dinger's equation in isolated systems - the quantum state evolves by the process of minimizing ``computational cost.''}

\maketitle
\flushbottom

\noindent

\section{Introduction}
Based on the intuition that the classical spacetime geometry encodes information theoretic properties of the dual quantum field theory (QFT)  in the context of gauge/gravity duality, many quantum information concepts have been applied to investigations of gravity theories. A notable example is the holographic entanglement entropy (EE) of a subregion in a QFT~\cite{Ryu:2006bv}.
Even though EE has played a crucial role in understanding the dual gravity, it turned out that EE is not enough~\cite{Susskind:2014moa}, in particular, when it comes to the interior of the black hole.  In the eternal AdS black hole, an Einstein-Rosen bridge (ERB) connecting two boundaries continues to grow for longer time scale even after thermalization. Because EE quickly saturates at the equilibrium, it cannot explain the growth of the ERB and   another quantum information concept, {\it complexity}, was introduced as a dual to the growth of ERB~\cite{Susskind:2014rva,Stanford:2014jda}.
To `geometrize'  the complexity of quantum states in the dual gravity theory, two conjectures were proposed: complexity-volume (CV) conjecture~\cite{Stanford:2014jda} and complexity-action (CA) conjecture~ \cite{Brown:2015bva}, which are called {\it holographic complexity}\footnote{Along these lines, there have been a lot of developments and we refer to some of them,~\cite{Cai:2016xho,Lehner:2016vdi,Chapman:2016hwi,Carmi:2016wjl,Reynolds:2016rvl,Kim:2017lrw,Carmi:2017jqz,Kim:2017qrq,Swingle:2017zcd}, for examples.}. See also Refs~\cite{Alishahiha:2015rta,Ben-Ami:2016qex,Couch:2016exn}.

However, note that the complexity in information theory is well-defined in {\it discrete} systems such as quantum circuits~\cite{Aaronson:2016vto}.  For example, the so-called {\it circuit complexity} is the minimal number of simple elementary gates required to approximate a target operator in quantum circuit.
On the contrary, holographic complexity is supposed to be dual to complexity in a QFT, a {\it continuous } system. Thus, there may be a mismatch in duality if we try to compare the holographic complexity with the results purely based on the intuition from circuit complexity and it is important to develop the theory of complexity in QFT.
Compared with much progress in holographic complexity, the precise meaning of the complexity in QFT is still not complete. In order to define complexity in QFT systematically, we start with the complexity of operator. The complexity between states will be obtained based on the complexity of operator. For the complexity of states we make a brief comment  in section \ref{sec8}  and refer to \cite{WIP} for more detail. Our strategy to define the complexity of operator is: (i) extract minimal and essential axioms for the complexity of operator from the circuit complexity and (ii) define the complexity in {\it continuous } QFT systems based on that  minimal axioms and smoothness (from continuity) iii) consider general symmetries of QFT to give constrain on the structure of complexity. It will turn out that these steps enable us to determine the complexity of the SU($n$) operators uniquely.

We want to  emphasize that {\it not all} properties of circuit complexity survive in the complexity in QFT.  The difference between discreteness and continuity makes some essential differences in properties of the complexity. For example, a few  basic concepts in ``circuit complexity'' (computational complexity), such as ``gates'',  are not well defined in general quantum QFT so they should be modified or abandoned. Thus, we will  keep only the most essential properties of the circuits complexity (which will be abstracted into the axioms \textbf{G1}-\textbf{G3} in the following section).  As another essential ingredient from QFT side, we will take advantages of basic symmetries of QFT, which may not be necessary  in the case of quantum circuits or computer science.
Because of the effect of this new inputs from QFT, some properties of the complexity in QFT we obtained may be incompatible with quantum circuits or qubit systems but they are more appropriate for QFT.


Our work is also inspired by a geometric approach by Nielsen et al. ~\cite{Nielsen1133,Nielsen:2006:GAQ:2011686.2011688,Dowling:2008:GQC:2016985.2016986}, where the discrete circuit complexity for a target  operator is identified with the minimal geodesic distance connecting the target operator and the identity in a certain Finsler geometry~\cite{038798948X,9810245319,xiaohuan2006an,asanov1985finsler}, which is just Riemannian geometry without the quadratic restriction. 
Recently, inspired by this geometric method, Refs.~\cite{Brown:2017jil,Jefferson:2017sdb,Yang:2017nfn,Kim:2017qrq,Khan:2018rzm} also investigated the complexity in QFT. However, in these studies, because the Finsler metric can be chosen arbitrarily, there is a shortcoming that the complexity depends on the choice of the metric. In this paper, we  show that, for SU($n$) operators, the Finsler metric and complexity are uniquely determined based on four general axioms (denoted by \textbf{G1-G4}) and the basic symmetries of quantum QFT.

\begin{figure}
  \centering
  \includegraphics[width=\textwidth]{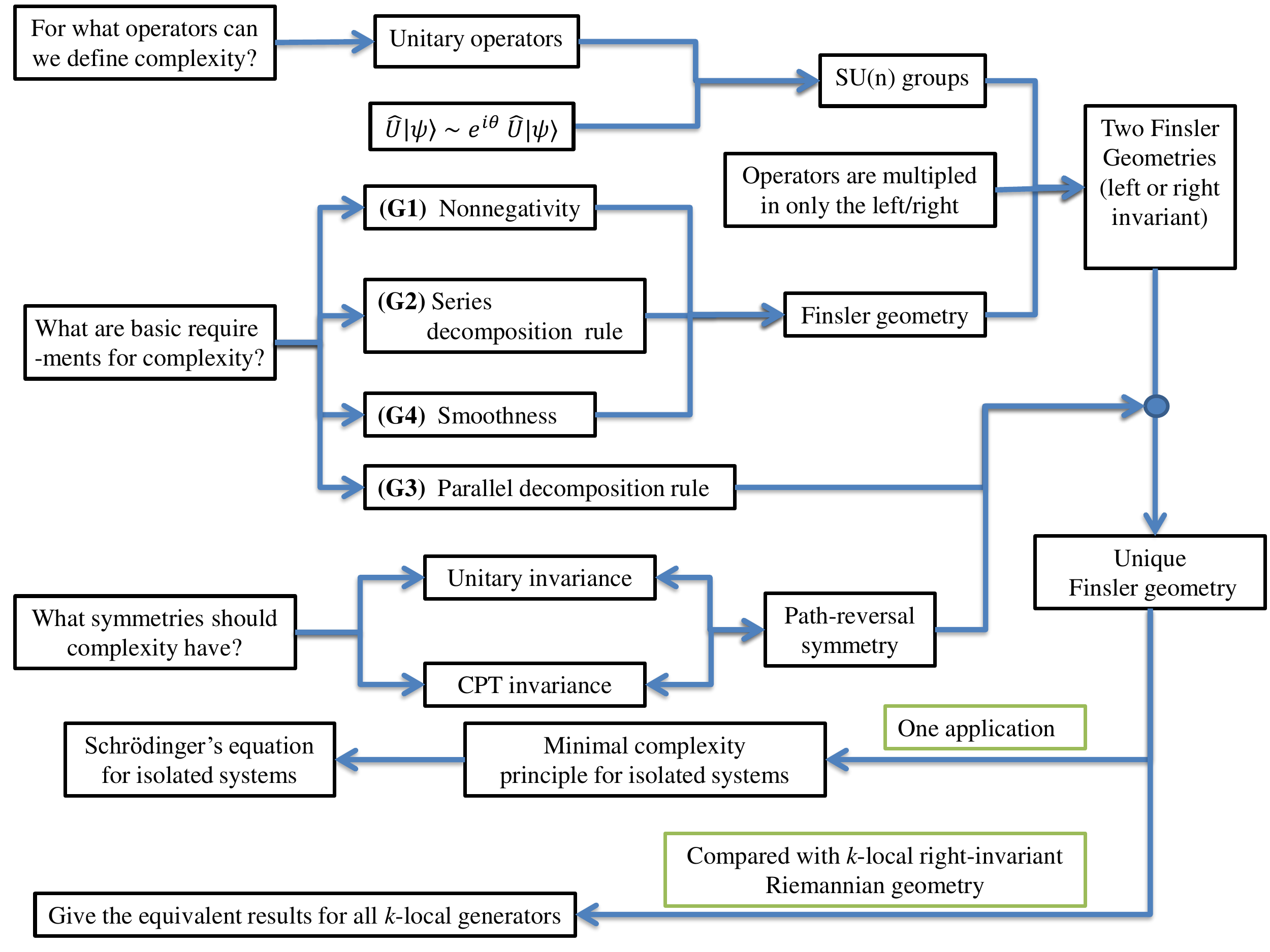}
  \caption{{The logical flows of this paper. By answering three basic questions at the far left side we show that the complexity geometry is determined by a unique  bi-invariant Finsler geometry.  As an application of our formalism, we show that the Schr\"{o}dinger's equation for isolated systems can be obtained from a ``minimal cost principle''.  Although the bi-invariant geometry looks very different from the right-invariant $k$-local Riemannian geometry proposed by Ref.~\cite{Brown:2017jil}, we will show that for all $k$-local operators two theory will give equivalent results.  }} \label{logic1}
\end{figure}

{In order to make our logic and claims clear we show a schematic map for the logic structure of this paper in Fig.~\ref{logic1}. We want to answer the following questions: (1) for what operators can we define complexity? (2) what are basic properties that complexity should satisfy? (3) for quantum field theory, what symmetries should appear in the complexity? (4) what can we obtain for complexity by the answers of above three questions? (5) what are the similarities and differences compared with previous works? }
%

Form the section perspective, this paper is organized as follows.
In section \ref{sec2}, we introduce minimal and basic concepts of the complexity and propose three axioms  \textbf{G1}-\textbf{G3} for the complexity of operators, which are inspired by the circuit complexity.
In section \ref{sec3}, we show how the Finsler metric arises from  \textbf{G1}-\textbf{G2}  and the smoothness of the complexity (\textbf{G4}).
In section \ref{sec4}, by using fundamental symmetry properties of QFT, we investigate constraints on the Finsler metric and the complexity. In particular we show that the Finsler metric is bi-invariant by several different approaches and is determined uniquely if we take the axiom \textbf{G3} into account.  We also compare our results with previous researches regarding bi-invariance. In section \ref{sec5} we derive the explicit form of the Finsler metric of the SU($n$) group.  Thanks to the bi-invariance, the geodesic in the Finsler space of SU($n$) group  (so the complexity) is easily computed.
 In section \ref{sec6}, as one application of the geodesic in the bi-invariance Fisnler metric, we propose a ``minimal cost principle'' as a new interpretation of the Schr\"{o}dinger's equation.
  In section \ref{sec7} we make  a comparison between our complexity and the complexity for $K$-qubit systems.
 In  section \ref{sec8} we conclude.

\section{Axioms for the complexity of  operators} \label{sec2}

\subsection{Why unitary operators?}

In order to make a good definition of the ``complexity of operator'' we first need to clarify what kind of ``operator'' we intend to deal with in this paper.

Intuitively, the complexity of operator measures how ``complex'' a physical process is. Thus, the operator should corresponds to a ``realizable'' physical process.  This concept can be formulated as follows.
An operator $\hat{O}$ is called $\varepsilon$-\textbf{realizable} if there is at least one experimental quantum process $\phi$ (for example, a quantum circuit) so that the following inequality holds for arbitrary two states $|\psi_1\rangle$ and $|\psi_2\rangle$
\begin{equation}\label{realizebleO}
  \left|\langle \psi_2|\hat{O}|\psi_1\rangle-\langle \psi_2|\phi(\psi_1)\rangle\right|^2\leq\varepsilon \,,
\end{equation}
with $\varepsilon > 0$.
Here $|\phi(\psi_1)\rangle$ is the output state of the quantum process $\phi$ for an input state $|\psi_1\rangle$.   The $\varepsilon$ is the tolerance when we use a $\phi$ to approximate (simulate) the target operator $\hat{O}$. Any physical system $\phi$ satisfying the inequality~\eqref{realizebleO} is called an $\varepsilon$-\textbf{realization} of operator $\hat{O}$ and denoted by $\phi_{\varepsilon,O}$. All $\varepsilon$-realizable operators form a set $\mathcal{O}_\varepsilon$.
{If an operator is $\varepsilon$-realizable for arbitrary positive $\varepsilon$, then we call it  a realizable operator. For example, the identity $\I$, which just keeps the input as the output, is one realizable operator.
All the realizable operators form the set $\mathcal{O}$.
Quantum system $\phi_{O}:=\lim_{\varepsilon\rightarrow0^+}\phi_{\varepsilon,O}$ is called a realization of operator $\hat{O}$.}

With the set of input states ($S_{\text{in}}$), the set of output states ($S_{\text{out}}$,) by realizable operators ($\hat{O} \in \mathcal{O} $) can be expressed as
\begin{equation}\label{SSU}
  S_{\text{out}} =\mathcal{O}S_{\text{in}} :=\{\hat{O}|\psi\rangle|\forall \hat{O}\in\mathcal{O}, \forall |\psi\rangle\in S_{\text{in}}\}\,.
\end{equation}
%
%
We assume
\begin{equation}\label{relSS}
  S_{\text{out}}\subseteq S_{\text{in}} \,,
\end{equation}
which makes it possible that the elements of output set can be used as new inputs. With the assumption~\eqref{relSS}, we can define the products of operators.
The product between elements in $\mathcal{O}$ can be defined as
\begin{equation}\label{defineO1O2}
  (\O_1\O_2)|\psi\rangle:=\O_1(\O_2|\psi\rangle),~\forall |\psi\rangle\,.
\end{equation}
By the definition of realizable operators, it can be shown that $\O_1\O_2$ is realizable if $\O_1$ and $\O_2$ are both realizable operators. Thus, $\mathcal{O}$ forms a monoid (semigroup with identity).

If we restrict physical processes to quantum mechanical processes, Eq.~\eqref{realizebleO} implies that realizable operators are all unitary rather than Hermitian.
In other words, our target is a property of the physical process rather than a direct observable.
{As quantum circuits are  quantum mechanical processes and  Solovay-Kitaev theorem~\cite{nielsen2010quantum} says that all the unitary operators can be approximated by some quantum circuits with any nonzero tolerance, we can conclude that the realizable operators set is the set of unitary operators. As unitary operators are invertible, the realizable operators set $\mathcal{O}$ forms a (finite dimensional or infinite dimensional) unitary group.\footnote{{Hermit operators, which correspond to observable quantities and are not unitary in general, cannot be approximated by quantum circuits if the tolerance is small enough. }}}

\subsection{Definitions and axioms}

{
Intuitively speaking, the circuit complexity (or computational complexity) of a target operator ( or computational task) is defined by the minimal number of required fundamental gates ( or fundamental steps) to simulate the target operator ( or finish the computational task).}
Based on this intuitive concept of the complexity  in quantum circuits and computations, we propose that the complexity defined in an arbitrary monoid $\mathcal{O}$ should satisfy the following three axioms. We denote a complexity of an operator $\hat{x}$ in an operators set $\mathcal{O}$ by $\mathcal{C}(\hat{x})$.

\begin{enumerate}
\item[\textbf{G1}] [{\textit{Nonnegativity}}]\\
$\forall \x\in\mathcal{O}$, ~$\mathcal{C}(\x)\geq0$ and the equality holds iff $\x $ is the identity.
\vspace{-0.2cm}
\item[\textbf{G2}] [{\textit{Series decomposition rule (triangle inequality)}}]\\
$\forall \x,\hat{y}\in\mathcal{O}$,  $\mathcal{C}(\x\hat{y}) \leq \mathcal{C}(\x)+\mathcal{C}(\hat{y})$. \vspace{-0.2cm}
\item[\textbf{G3}] [{\textit{Parallel decomposition rule}}] \\
$\forall  (\x_1 ,\hat{x}_2)  \in \mathcal{N}= \mathcal{O}_1 \times \mathcal{O}_2 \subseteq  \mathcal{O}$,  $\mathcal{C}\big((\hat{x}_1,\hat{x}_2)\big)=\mathcal{C}\big((\hat{x}_1,\I_2)\big)+\mathcal{C}\big((\I_1,\hat{x}_2)\big)$.
\end{enumerate}%
Here, in \textbf{G2}, it is possible that the operator $\hat{x}\hat{y}$ is decomposed in different ways, say $\hat{x}' \hat{y}'$. In this case, \textbf{G2} can read also as $\mathcal{C}(\x\hat{y})  = \mathcal{C}(\x'\hat{y}') \leq \mathcal{C}(\x')+\mathcal{C}(\hat{y}')$. In \textbf{G3}, we consider the case that there is a sub-monoid $\mathcal{N} \subseteq\mathcal{O}$  which can be decomposed into the Cartesian product of two monoids, i.e., $\mathcal{N}= \mathcal{O}_1 \times \mathcal{O}_2 $. $\I_1$ and $\I_2$ are the identities of $\mathcal{O}_1$ and $\mathcal{O}_2$.  The Cartesian product of two monoids implies that $(\x_1,\x_2 )(\hat{y}_1,\hat{y}_2)=(\x_1\hat{y}_1, \x_2\hat{y}_2)$  for arbitrary  $(\x_1,\x_2),  (\hat{y}_1,\hat{y}_2)\in\mathcal{N}$.

The axiom \textbf{G1} is obvious by definition. We call the axiom \textbf{G2} ``series decomposition rule'' because the decomposition of the operator $\O=\x\hat{y}$ to $\x$ and $\hat{y}$ is similar to the decomposition of a big circuit into a series of small circuits. Reversely, the `product' of two operators corresponds to a serial connection of two circuits. The axiom \textbf{G2} answers a basic question: what is the relationship between the complexities of two operators and the complexity of their products?
Because the complexity is a kind of ``minimal'', we require the inequality in \textbf{G2}.\footnote{The reason for axiom \textbf{G2} has been explained clearly also in the Nielson' works~\cite{Nielsen1133,Nielsen:2006:GAQ:2011686.2011688,Dowling:2008:GQC:2016985.2016986}.} This $\textbf{G2}$  will lead to the familiar ``triangle inequality'' in the concept of distance (see \textbf{F3} in the Sec.~\ref{sec3}) so it is also called ``triangle inequality''.

\begin{figure}
  \centering
  \includegraphics[width=.5\textwidth]{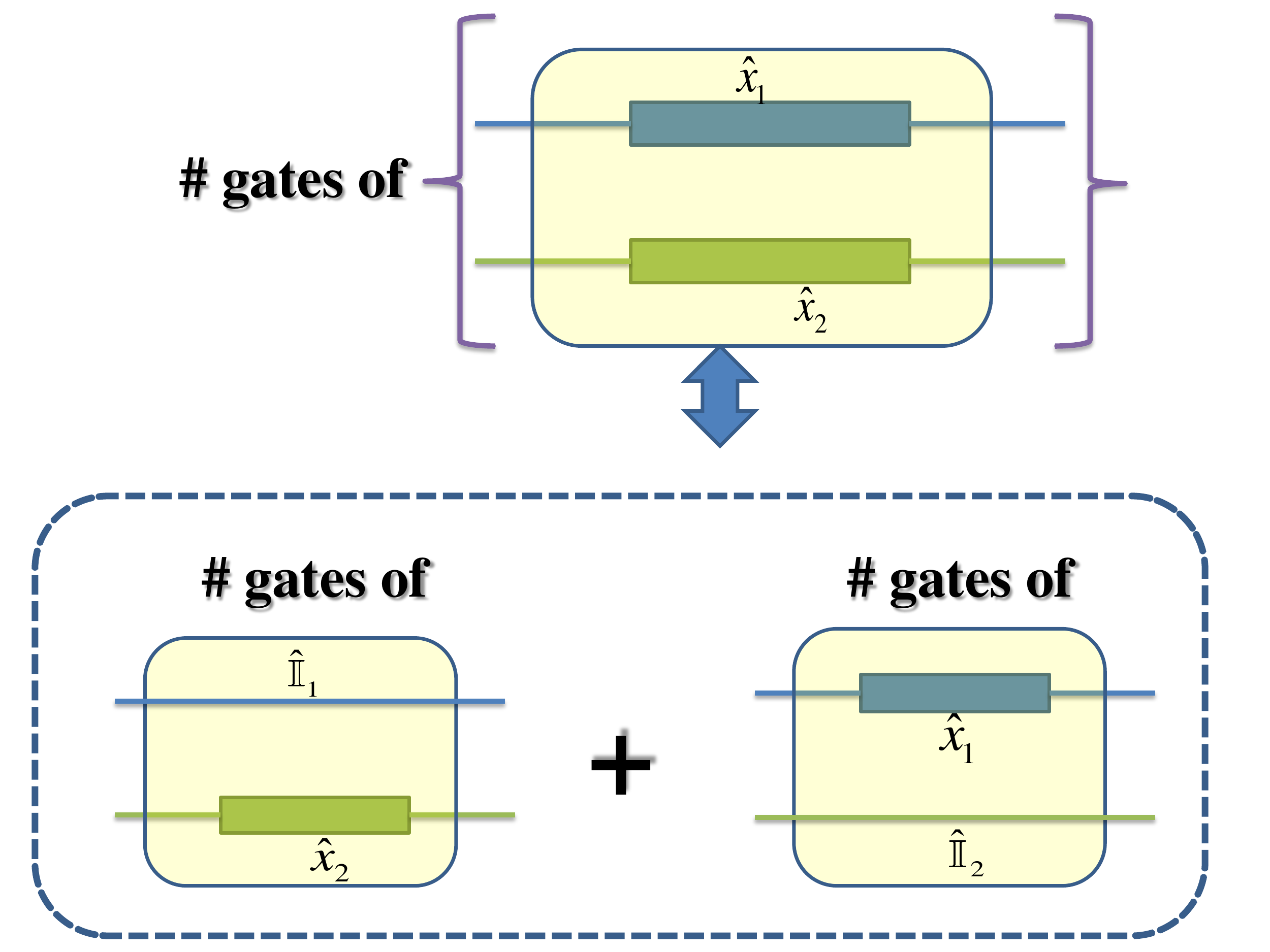}
  \caption{Schematic diagram for the complexity of the Cartesian product and parallel decomposition rule. As two operators $\hat{x}_1$ and $\hat{x}_2$ are simulated independently, the minimally required gates for $(\hat{x}_1,\hat{x}_2)$ is the sum of the minimally required gates for $\hat{x}_1$ and $\hat{x}_2$. Thus, we have $\mathcal{C}((\hat{x}_1,\hat{x}_2))=\mathcal{C}((\hat{x}_1,\I_2))+\mathcal{C}((\I_1,\hat{x}_2))$. }
   \label{Cdproduct}
\end{figure}

In contrast to \textbf{G2} (series decomposition rule), we call the axiom \textbf{G3} ``parallel decomposition rule'', \kyr{which is chosen as one of the most basic axioms in defining complexity for the first time in this paper.\footnote{\kyr{In some cases, the parallel decomposition may be impossible. However, in local computations/gates, the parallel decomposition is permitted, for example see Ref.~\cite{Vanchurin:2017qii}. Our axiom \textbf{G3} does not talk about the possibility of the parallel decomposition, but says what will happen to the complexity if the parallel decomposition is permitted.} }}
 It comes from the following fundamental question: if an operator (task) $\O$ contains two \textit{totally independent} sub-operators (sub-tasks) $\x_1$ and $\x_2$, what should be the relationship between the total complexity  and the complexities of  two sub-operators (sub-tasks)? Here, the  \textit{totally independent} means that: (a) $\O$ accepts two inputs and yields two outputs through $\x_1$ and $\x_2$, and (b) the inputs for $\x_1$ (or $\x_2$) will never affect the outputs of $\x_2$ (or $x_1$). See Fig. \ref{Cdproduct} for this explanation.

Mathematically, the construction of a bigger operator $\O$ by $\x_1$ and $\x_2$ under two requirements (a) and (b) corresponds to the \textit{Cartesian product} denoted by $\O=(\x_1,\x_2)$.  Note that the Cartesian product of two monoids does not correspond to the tensor product in a linear representation (i.e., a matrix representation). Instead, it corresponds to the direct sum. For example, if matrixes $M_1$ and $M_2$ are two representations of operators $\x_1$ and $\x_2$, then the representation of their Cartesian product $\O$ is $M_1\oplus M_2$, which is neither $M_1\otimes M_2$ nor  $M_1M_2$.

In the language of computer science,  this  ``\textit{totally independent}'' just means that one task contains two independent \textit{parallel} tasks. Thus, the axiom \textbf{G3} tries to answer the following question: if a task contains two parallel sub-tasks, what should be the relationship between the total complexity and the complexities of such sub-tasks?  In term of mathematical language, it amounts to asking: what should be the relationship between $\mathcal{C}\big((\hat{x}_1,\hat{x}_2)\big)$, $\mathcal{C}\big((\hat{x}_1,\I_2)\big)$ and $\mathcal{C}\big((\I_1,\hat{x}_2)\big)$?

\textbf{G3} answers this question by requiring that the complexity of two parallel tasks is the sum of their complexities, which is very natural. See Fig.~\ref{Cdproduct} for a schematic explanation. In matrix representation, \textbf{G3} says, for an operator $M=M_1\oplus M_1$, $\C(M_1\oplus M_1)=\C(M_1)+\C(M_2)$. It can be generalized to the direct sum of many operators:  for {a finite number of matrixes} $M_1, M_2, \cdots, M_k$, we have
\begin{equation}\label{eqforG30}
   \text{Parallel decomposition rule }\textbf{G3: }~~\C\left(\bigoplus_{i=1}^k M_i\right)=\sum_{i=1}^k\C(M_i)\,.
\end{equation}

One may worry about the self-consistency between \textbf{G2} and \textbf{G3} and argue that we can only require $\mathcal{C}((\hat{O}_1, \hat{O}_2))\leq\mathcal{C}(\hat{O}_1)+\mathcal{C}(\hat{O}_2)$, as there may be other operators $\{\O_a, \O_a', \O_b, \O_b'\}$ to satisfiy $(\O_a,\hat{O}_a') (\hat{O}_b,\O_b')=(\hat{O}_1, \hat{O}_2)$ but the total gates is less than $\C(\O_1)+\C(\O_2)$. However, this is impossible.
One can see that the sum of the minimal gates of $\{\O_a, \O_a', \O_b, \O_b'\}$ is $\C(\O_a)+\C(\O_a')+\C(\O_b)+\C(\O_b')$. But according to the fact that $\O_a\O_a'=\O_1$ and $\O_b\O_b'=\O_2$, we find that
$$\C(\O_a)+\C(\O_a')+\C(\O_b)+\C(\O_b')\geq\C(\O_1)+\C(\O_2) .$$
Thus, $\mathcal{C}(\hat{O}_1)+\mathcal{C}(\hat{O}_2)$ is the minimal gates to obtain $(\hat{O}_1, \hat{O}_2)$.

The axioms \textbf{G1}-\textbf{G3} are satisfied by both circuit complexity and computational complexity.  We have expressed the abstract concepts extracted from circuit complexity and computational complexity in terms of mathematical language and will take them as three basic requirements to define complexity also in other systems. The axiom \textbf{G1} and \textbf{G2} can be satisfied by Nielson's original works Refs.~\cite{Nielsen1133,Nielsen:2006:GAQ:2011686.2011688,Dowling:2008:GQC:2016985.2016986} and recent other approaches to complexity such as Refs.~\cite{Brown:2017jil,Jefferson:2017sdb,Yang:2017nfn,Kim:2017qrq,Khan:2018rzm}. However, these works did not take into account the question related to \textbf{G3} and broke the requirement in  axiom \textbf{G3} in general.
From the viewpoint of quantum circuits (or computer science), series circuits (or tasks ) and parallel circuits (or tasks) are two fundamental manners to decompose a bigger circuits (or tasks) into smaller ones. Thus, the axioms \textbf{G3} should be as important as \textbf{G2}. In this paper, we propose the concept of  \textbf{G3}  for the first time and show that it plays a crucial role in determining the form of the complexity of SU($n$) operators. We may be able to modify \textbf{G3} in somewhat {\it unnatural} way, which will lead us to another form of the Finsler metric similar to \eqref{RFmetric}. This  point will be clarified in more detail in Ref.~\cite{Yang:2018tpo}.


\section{Emergence of the Finsler structure from the axioms for the complexity}  \label{sec3}

In this section, we show that the Finsler metric arises from the minimal and general axioms for the complexity  \textbf{G1}-\textbf{G3}  and the smoothness of the complexity.  {From here, the group element may represent either an abstract object or a faithful representation, which  will be understood by context.  }

In section \ref{sec2} we have shown that the realizable operators are unitary operators, so the question now becomes how to define the complexity for unitary operators. As the unitary operators $\hat{U}$ and $e^{i\theta} \hat{U}$ (with $\theta\in(0,2\pi)$)  produce equivalent quantum states, the complexity of $\hat{U}$ and $e^{i\theta} \hat{U}$ should be the same. Thus it is enough to study the complexity for special unitary groups, SU($n$) groups. Ultimately, our aim is to investigate the complexity for operators in quantum field theory, of which Hilbert space is infinite dimensional, so we have to deal with the infinite dimensional special unitary groups. However, they involve infinite dimensional manifolds and have not been well-studied even in mathematics so far. As an intermediate step, in this paper, we will first present our whole theory for finite dimensional cases and assume that the results can be generalized into infinite dimensional cases by some suitable limiting procedures. The subtle aspects between finite and infinite dimensional Lie groups are now under investigation~\cite{WIP}.

\begin{figure}
 \centering
  \includegraphics[width=.4\textwidth]{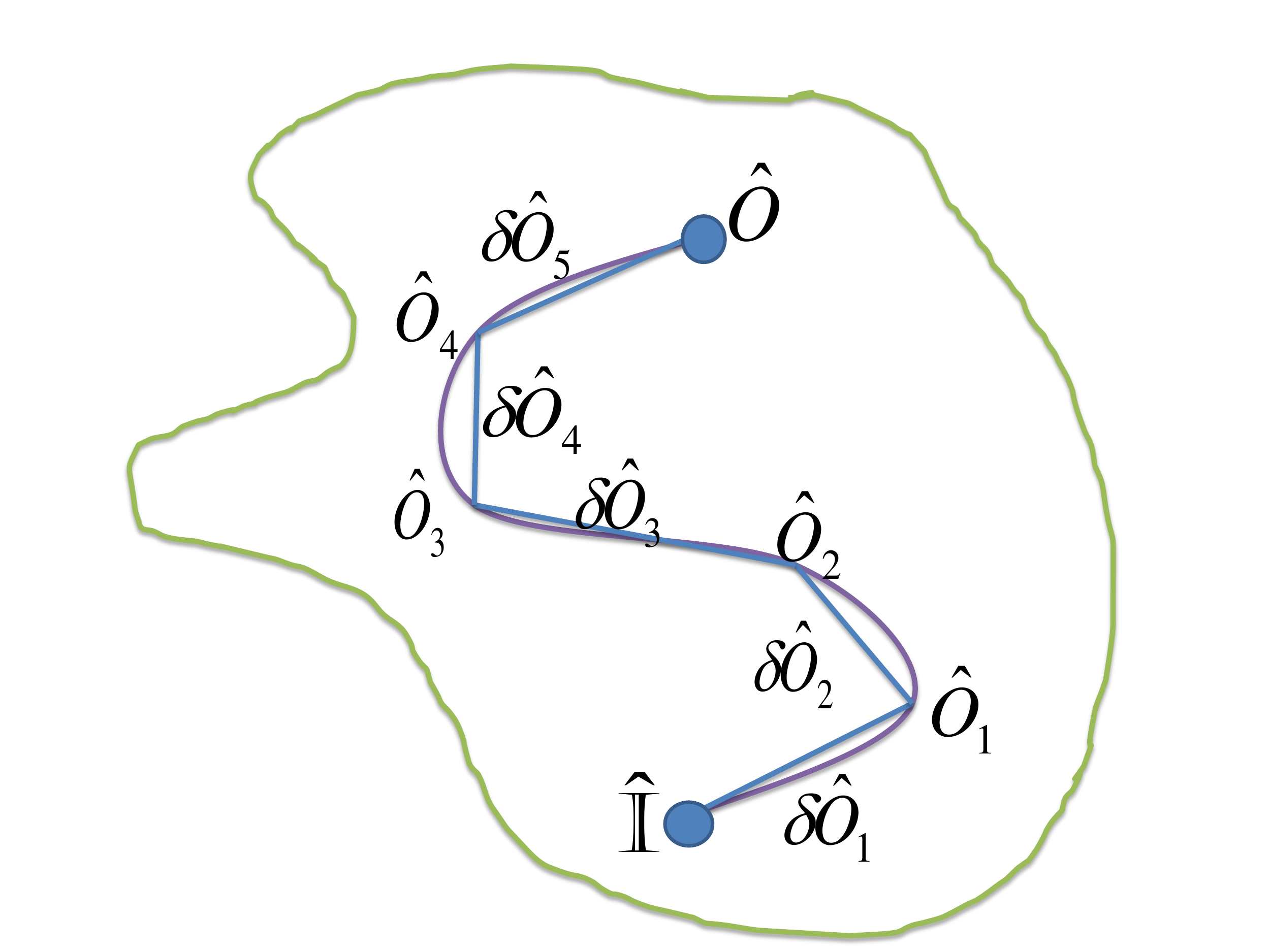}
  \caption{A curve $c(s)$ connects the identity and a particular operator $\hat{O}$ with the endpoints $c(0)=\I$ and $c(1)=\hat{O}$. This curve can be approximated by a discrete form. Every endpoint is also an operator, which is labeled by $\hat{O}_n$
} \label{expalinF1}
\end{figure}

For a given operator $\hat{O}\in\text{SU}(n)$, as $\text{SU}(n)$ is connected, there is a curve $c(s)$ connecting $\hat{O}$ and identity $\I$. The curve  may be parameterized by  $s$ with $c(0)=\I$ and $c(1)=\hat{O}$. {See Fig.~\ref{expalinF1}.}
The tangent of the curve, $ \dot{c}(s)$, is assumed to be given by a right generator $H_r(s)$ or a left generator $H_l(s)$:
\begin{equation}\label{rightH1}
\dot{c}(s) =H_r(s)c(s)  \ \ \  \mathrm{or} \ \  \  \dot{c}(s) =c(s)H_l(s)  \,.  
\end{equation}
%
%
This curve can be approximated by discrete forms:
%
{
\begin{eqnarray}
  \hat{O}_n&&=c(s_n) \\
   &&= \delta\hat{O}_n^{(r)}\hat{O}_{n-1} \label{discreteR}\\
   && =\hat{O}_{n-1} \delta\hat{O}_n^{(l)}\,, \label{discreteL} 
\end{eqnarray}
where  $s_n=n/N$, $n=1,2,3,\cdots, N$, $\hat{O}_{0}=\I$  and $\delta\hat{O}_{n}^{(\alpha)}=\exp[H_\alpha(s_{n})\delta s]$ with $\alpha=$ $r$ or $l$ and $\delta s=1/N$.} In general, the two generators $H_r(s)$ and $H_l(s)$ at the same point of the same curve can be different, i.e., $H_r(s)\neq H_l(s)$. In fact, from Eq. \eqref{rightH1}, we see $H_r(s)$ is one adjoint transformation of $H_l(s)$,
\begin{equation}\label{relHrHl}
  H_r(s)=c(s) H_l(s)c(s)^{-1}\,.
\end{equation}

\begin{figure}
  \centering
  \includegraphics[width=.5\textwidth]{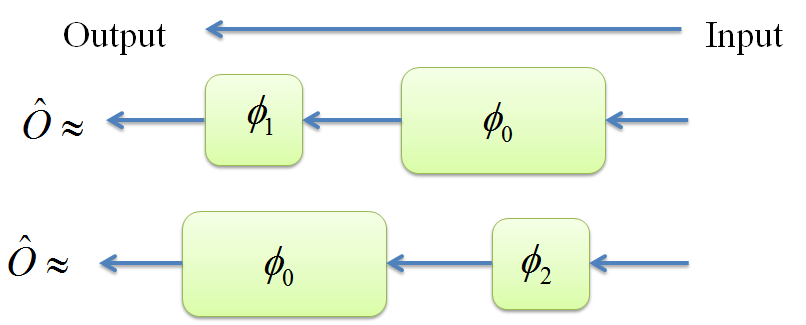}
  \caption{Schematic diagram for two different generators in quantum circuits. To obtain the some target operator $\O$ from the quantum circuit $\phi_0$, we have two different ways to add new circuits.}
   \label{RLphi1}
\end{figure}

The availability of two different generators can be understood also by a quantum circuit approximation to an operator, say $\hat{O}$.  As shown in Fig.~\ref{RLphi1}, if a quantum circuit  $\phi_0$  is given, the operator $\hat{O}$ can be constructed in two ways: i) by adding a new quantum circuit $\phi_1$ after the output of $\phi_0$ (corresponding to Eq. \eqref{discreteR}) or ii) by adding a new quantum circuit $\phi_2$ before the input of $\phi_0$ (corresponding to Eq. \eqref{discreteL}). 
The previous works such as Refs.~\cite{Nielsen1133,Nielsen:2006:GAQ:2011686.2011688,Dowling:2008:GQC:2016985.2016986,Brown:2017jil} assumed that the new operators/circuits could appear only after the output side of original operators/circuits, which corresponds to Eq. \eqref{discreteR}. This  is one mathematically allowed choice but there is no a priori or a physical reason for that particular choice. Eq. \eqref{discreteL} should be equally acceptable.

The axioms \textbf{G1}-\textbf{G3} are suitable for arbitrary monoid, both discrete and continuous ones.  Now SU($n$) group is a manifold, it is natural to expect that the complexity on it is a smooth function. In fact, it turns out to be enough to assume a weaker form
\begin{enumerate}
\item [\textbf{G4}] [\textit{Smoothness}] The complexity of any infinitesimal operator in SU($n$), $\delta \hat{O}^{(\alpha)} = \exp (H_\alpha \delta s)$, is a smooth function of  {\it only}  $H_\alpha \ne 0$ and $\delta s \ge 0$, i.e.,
\begin{equation} \label{compdel}
\mathcal{C} (\delta \hat{O}^{(\alpha)}) = \mathcal{C}(\I) + \tilde{F}(H_\alpha) \delta s + \mathcal{O} (\delta s^2) \,,
\end{equation}
where $\tilde{F}(H_\alpha) := \partial_{\delta s} \mathcal{C}  (\delta \hat{O}^{(\alpha)})|_{\delta s =0}  $ and  $\mathcal{C}(\I) = 0$ by \textbf{G1}.
\end{enumerate}
which is our forth axiom.  Notice that   $\C(\delta\O^{(r)})=\C(\delta\O^{(l)})$ if $\delta\O^{(r)}=\delta\O^{(l)}$, which  implies that an infinitesimal operator will give the same contribution to the total complexity when it is added  to the left-side or right-side.\footnote{{This condition will be relaxed in our upcoming work~\cite{Yang:2018tpo}, where we will allow different contributions to the total complexity when $\delta\O$ is added to the left-side or right-side. }} Thus, the index $\alpha$ is in fact not necessary in this case, but we keep it for notational consistency.   

Let us define the {\it cost} ($L_\alpha[c]$) of a particular curve $c$, constructed by only $\delta\hat{O}_n^{(r)}$ or only $\delta\hat{O}_n^{(l)}$, as
\begin{equation} \label{Lalpha}
L_\alpha[c] := \sum_{i=1}^N\mathcal{C} (\delta \hat{O}_i^{(\alpha)})  \xrightarrow{N \rightarrow \infty} \int_0^1 \tilde{F} (H_\alpha (s)) \td s \,.
\end{equation}
Geometrically, it is the length of the particular curve and $\tilde{F}  \td s  $ looks like a line element in some geometry. Thus, the natural question will be what kind of geometry is allowed for complexity? We will show that it is Finsler geometry, which emerges naturally from our axioms for the complexity.

First, we can prove that $\tilde{F}$ satisfies three properties:
\begin{enumerate}
\item[\textbf{F1}](Nonnegativity) $\tilde{F}(H_\alpha)\geq0$ and $\tilde{F}(H_\alpha)=0$ iff $H_\alpha=0$ \vspace{-0.2cm}
\item[\textbf{F2}](Positive homogeneity) $\forall\lambda\in\mathbb{R}^+$, $\tilde{F}(\lambda H_\alpha)=\lambda\tilde{F}(H_\alpha)$ \vspace{-0.2cm}
\item[\textbf{F3}](Triangle inequality)  $\tilde{F}(H_{\alpha,1})+\tilde{F}(H_{\alpha,2})\geq\tilde{F}(H_1+H_2)$
\end{enumerate}
only by using \textbf{G1}, \textbf{G2} and  \textbf{G4}!
(see  appendix~\ref{app1} for a proof.)
Note that \textbf{F1}-\textbf{F3} may describe some suitable properties that the concept of the  `norm' of vectors in a vector space should satisfy.  In our case, the vector space is the Lie algebra (the tangent space at the identity) and the generators ($H_\alpha$) of the algebra are vectors. Indeed, the `norm' satisfying the properties \textbf{F1}-\textbf{F3} is called a \textit{Minkowski norm} in mathematical jargon.
Once we know $H_\alpha(s)$ we can compute the length of the line element by a Minkowski norm $\F$. (At this stage, we don't know the explicit form of the Minkowski norm, but we will determine it later.)

For a given $\F$, we have two different natural ways to extend the Minkowski norm $\tilde{F}$ at the identity to every point on the base manifold via arbitrary curves.
\begin{equation} \label{rlinvar}
F_r(c, \dot{c}) := \tilde{F} (H_r)=\F(\dot{c} c^{-1})\,, \  \ \mathrm{or} \ \ F_l(c, \dot{c}) := \tilde{F} (H_l)=\F(c^{-1}\dot{c}) \,.
\end{equation}
where we introduce a new notation `$F_\alpha(c, \dot{c})$', a standard notation for {\it Finsler metric} in mathematics. The introduction of `$F_\alpha(c, \dot{c})$' is justified because the Finsler metric is nothing but a Minkowskia norm defined at all points on the base manifold and Eq. \eqref{rlinvar} explains how to assign the Minkowskia norm to all the other points. We refer to Refs.~\cite{038798948X,9810245319,xiaohuan2006an,asanov1985finsler} for an introduction to  Minkowski norm and the Finsler geometry.\footnote{{Strictly speaking, $\F$ and $F_\alpha$ are the Minkowski norm and the Finsler metric respectively  if we make a further requirement that, in \textbf{F3}, the equality holds only when $H_1$ and $H_2$ are linearly dependent. However, this subtle mismatch is not important physically.}} \kyr{A brief  introduction to the Finsler geometry can be found in appendix~\ref{introF}.}

There is an invariant property in the Finsler metrics. $F_r(c, \dot{c})$ is {\it right-invariant} {because} {$H_r$} is invariant under the right-translation $c \rightarrow c \hat{x}$ for $\forall \hat{x} \in $ SU($n$). Similarly $F_l(c, \dot{c})$ is {\it left-invariant} {because} {$H_l$} is invariant under the left-translation $c \rightarrow  \hat{x} c$ for $\forall \hat{x} \in $ SU($n$).
If there is no further restriction on $F_\alpha$, there are at least two natural Finsler geometries, $F_r$ or $F_l$, which may give different cost or length.

%

Finally, the left or right complexity of an operator ($\mathcal{C}_\alpha(\hat{O})$) is identified with the minimal length (or minimal cost) of the curves connecting $\I$ and $\hat{O}$:
\begin{equation}\label{defcomF1}
  \mathcal{C}_\alpha(\hat{O}) :=\min \{L_\alpha[c]|~\forall c(s),~c(0)=\I,~c(1)=\hat{O}\}\,.
\end{equation}
We see that, even if we know the complexity of every infinitesimal operator (Eq. \eqref{compdel}), we have at least two different ways (left or right-way) to define the complexity of an operator and there is no a priori preferred choice among them. In order for the complexity of an operator to be a well-defined physical observable, this mathematical ambiguity should disappear naturally by some suitable physical considerations. In the following section we will show how this ambiguity is removed.

\section{Symmetries of the complexity inherited from QFT symmetries}  \label{sec4}
In the previous section, we have shown that the complexity can be computed by the minimal length of curves in {\it Finsler }geometry. We want to emphasize again that in our work the Finsler structure is {\it not assumed}, but it has been {\it derived} based on \textbf{G1}, \textbf{G2} and  \textbf{G4}. This is a  novel feature of our work compared to other works dealing with the Finsler geometry.

However, apart from the defining properties of the Finsler metric \textbf{F1}-\textbf{F3}, we don't know anything on $\tilde{F}(H_\alpha)$ so far. In this section, we will show there are constraints on $\tilde{F}(H_\alpha)$ if we take into account some symmetries of QFT. This is another important novel feature of our work compared to others. From here, we do not rely on properties of discrete systems or circuit models, which may be incompatible with QFT so may mislead us. We will directly deal with QFT and its symmetry properties and see what kind of constraints we can impose on $\tilde{F}(H_\alpha)$.

Note that such symmetry considerations are not necessary if we use ``complexity'' as a purely mathematical tool, for example, to study the ``NP-completeness'' and to analyze how complex an algorithm or a quantum circuit is. However, when we use the complexity to study real physical processes and try to treat the complexity as a basic physical variable hiding in physical phenomena, symmetries relevant to physical phenomena will be a necessary requirement.

In subsection \ref{sec41}, by requiring unitary invariance for complexity  we find
\begin{equation} \label{const11}
\text{[Independence of left/right generators]} \quad  \tilde{F}(H_l)=\tilde{F}(H_r) \,.
\end{equation}
It  means that the complexity does not depend on our choice of $H_r$ or $H_l$. Therefore, we call this property `Independence of left/right generators' of $\tilde{F}$.  Recall that for a given curve, we may have two metrics, either $F_r(c,\dot{c})=\F(H_r)$ or $F_l(c,\dot{c})=\F(H_l)$. It is an inherent ambiguity mathematically but this ambiguity can be removed by imposing physical condition, unitary invariance.
To support our result \eqref{const11} we will present three more arguments in  subsections~\ref{sec42} and appendix~\ref{othermetric}. Note that the constraint \eqref{const11} also implies the Finsler geometry is {\it bi-invariant}, meaning both right and left invariant.

 In subsection \ref{CPTinva}, by requiring the CPT symmetry\footnote{The CPT symmetry is a theorem for local relativistic quantum field theories in Minkowski space-time. Here, C means `charge conjugation', P `parity transformation' (`space inversion'), and T `time reversal'. This theorem states that  the local  Lorentz quantum field theories are invariant under the combined transformations of C, P, and T.},   we obtain
\begin{equation} \label{const22}
\text{[reversibility]}\quad  \F(H_\alpha) = \F(-H_\alpha) \,.
\end{equation}
We call this property `reversibility' of $\tilde{F}$ following the mathematical literature, for example, ~\cite{xiaohuan2006an}. In appendix~\ref{othermetric}, we will provide two more methods to support Eq. \eqref{const22}. Geometrically speaking, for a given path connecting $A$ and $B$, it is the constraint Eq.~\eqref{const22} that gives the same length when we go from $A$ to $B$ and from $B$ to $A$.

\subsection{Independence of left/right generators from unitary invariance} \label{sec41}
In this subsection we  consider the effect of the unitary invariance of the quantum field theory on the Finsler metric, cost, and complexity.
Let  us consider an arbitrary quantum field $\Phi$ with a Hilbert space $\mathcal{H}$ and a vacuum $|\Omega\rangle$, which are collectively denoted by $\{\Phi,\mathcal{H}, |\Omega\rangle\}$. Its unitary partner is  $\tilde{\Phi}(\vec{x},t):=\hat{U}\Phi(\vec{x},t)\hat{U}^{\dagger}$, $\tilde{\mathcal{H}}:=\{\hat{U}|\psi\, \rangle|\, \forall |\psi\rangle\in\mathcal{H}\}$ and $|\tilde{\Omega}\rangle:=\hat{U}|\Omega\rangle$, which are denoted by $\{\tilde{\Phi},\tilde{\mathcal{H}},|\tilde{\Omega}\rangle\}$.

In the Heisenberg picture, the dynamic of the quantum field $\Phi$ is governed by a time evolution operator $c(t)$:
\begin{equation}\label{dynamicPhi1}
  \Phi(\vec{x},t)=c(t)^\dagger \Phi(\vec{x},0)c(t)\,.
\end{equation}
%
The time evolutions of its unitary partner  $\tilde{\Phi}$  is
\begin{equation}\label{unitrypartner1}
\begin{split}
  \tilde{\Phi}(\vec{x},t)&=\hat{U}c(t)^\dagger\hat{U}^{\dagger}\hat{U}\Phi(\vec{x},0)\hat{U}^\dagger\hat{U}c(t)\hat{U}^{\dagger}\\
  &=\tilde{c}(t)^\dagger\tilde{\Phi}(\vec{x},0)\tilde{c}(t)\,,
  \end{split}
\end{equation}
where
\begin{equation}\label{dynamicPhi11}
  \tilde{c}(t):=\hat{U}c(t)\hat{U}^{\dagger}  \,,
\end{equation}
%
%
%
so the evolution of the unitary partner $\tilde{\Phi}$ is given by $\tilde{c}(t)$.
On the other hand, we cannot distinguish $\{\Phi,\mathcal{H},|\Omega\rangle\}$  and its unitary partner $\{\tilde{\Phi},\tilde{\mathcal{H}},|\tilde{\Omega}\rangle\}$ in the sense that any physical experiment will be invariant under the transformation $\{\Phi,\mathcal{H},|\Omega\rangle\}\rightarrow\{\tilde{\Phi},\tilde{\mathcal{H}},|\tilde{\Omega}\rangle\}$.
We will call this invariance ``unitary-invariance''.
%
%
Thus, it is  natural to expect that the cost cannot distinguish them either,  i.e.\footnote{\kyr{It seems that the complexity is invariant under the unitary transformation is weaker than the requirement that the cost function is invariant under the unitary transformation. However, it can be shown that they are equivalent as shown in appendix ~\ref{lastapp} }}
\begin{equation}\label{adjointcs0}
  L_\alpha[c]=L_\alpha[\hat{U}c(t)\hat{U}^{\dagger}]\,,~~\forall \hat{U}\in\text{SU($n$)}   \,.
\end{equation}

\begin{figure}[]
  \centering
  \includegraphics[width=.5\textwidth]{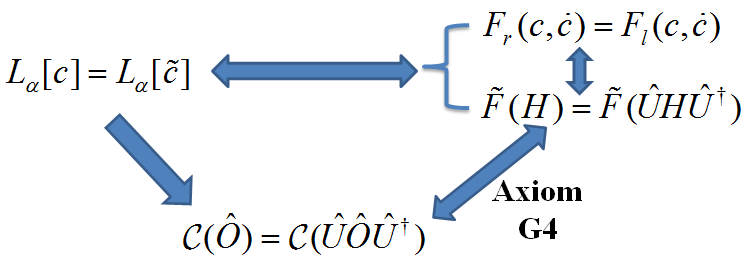}
  \caption{{Equivalences between  i) the unitary-invariance of QFT, ii) the independence of the Finsler metric on the left/right generators, and iii) the adjoint invariance of the complexity}}
   \label{triangle1}
\end{figure}

To extract a constraint on $\tilde{F}$ imposed by Eq. \eqref{adjointcs0}, it is enough to consider a special curve generated by an arbitrary constant generator $H_\alpha$
\begin{equation} \label{constH11}
c(t)=\exp( H_\alpha t) \,,
\end{equation}
with $t\in[0,1]$.
By the definition of the cost, Eq. \eqref{Lalpha}, we have
\begin{equation}\label{arugueA1s211}
  \int_0^1\F(H_\alpha)\td t=\int_0^1\F(\hat{U}H_\alpha \hat{U}^\dagger)\td t\,,
\end{equation}
which implies
%
\begin{equation}\label{adjinva1}
\text{[adjoint\ invariance]}\quad \tilde{F}(H_\alpha) = \tilde{F}(\hat{U}H_\alpha \hat{U}^\dagger)\,, \qquad  \forall U \in \text{SU($n$)} \,,
\end{equation}
As $H_r$ is just one adjoint transformation of $H_l$ (see Eq.~\eqref{relHrHl}),  it follows that
\begin{equation}\label{bi-inv}
\F(H_r)=\F(H_l) \quad \text{or} \quad F_r(c,\dot{c})=F_l(c,\dot{c})\,,
\end{equation}
where Eq.~\eqref{rlinvar} is used. It means that the left generator and the right generator give the same complexity.  Although we have the freedom to choose the left or right generator, the complexity will be independent of our choice.  In other words, if we know the complexity for every infinitesimal operator (Eq. \eqref{compdel}), then we have a unique value of the complexity in spite of the inherent ambiguity due to the availability of the left and right generators.    In Fig.~\ref{triangle1}, we summarize the relation between the  constraints  on the Finlser metric, cost, and complexity. One important consequence of Eq. \eqref{bi-inv} is that the Finsler geometry is {\it bi-invariant}, which means both right and left invariant. This property will be very useful when we determine the geodesic in the geometry in section \ref{sec52}.
%

One may argue that the complexity may not be directly observable and it is possible that $c(s)$ and $\tilde{c}(s)$ give different complexity. If that happens in some framework of computing the complexity, in our opinion,  there must be some gauge freedom in the definition of the complexity in the framework, for the complexity still to be a physical object. Thus, we will be able to make a suitable gauge fixing or a redefinition of the complexity so that this ``new complexity'' is physical and satisfies Eq.~\eqref{adjointcs0}.
%
%

%
\subsection{Comparison between SU($n$) groups and qubit systems} \label{sec42}

\begin{figure}
  \centering
  \includegraphics[width=.75\textwidth]{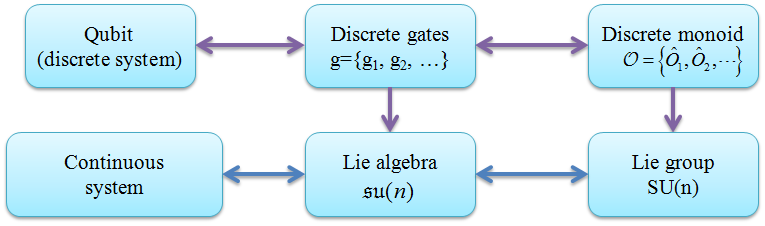}
  \caption{Schematic diagram for the relation and difference between qubit systems and SU($n$) groups. }
   \label{relqubitSUn}
\end{figure}

In order to clarify why the adjoint invariance of the complexity is natural, which may not be the case in discrete systems, we make a comparison between SU($n$) groups and qubit systems  in Fig.~\ref{relqubitSUn}.
For qubit systems, the operators set forms a countable monoid $\mathcal{O}$ and can be obtained from a countable fundemental gates set $g$. The complexity of any operator in $\mathcal{O}$ is given by the minimal gates number when we use the gates in $g$ to form the target operator. {For SU($n$) groups}, the operators set forms an SU($n$) Lie group and the fundamental gates are replaced by infinitesimal operators, which form the Lie algebra $\mathfrak{su}(n)$.

For qubit systems, suppose that the complexity measured by gates set $g$ is $\C_{\alpha}(g;\O)$ (here $\alpha=r,l$). If we make a ``global'' {unitary} transformation on the operators set and gates set together, i.e., $\tilde{\mathcal{O}}:=\hat{U} \mathcal{O}\hat{U}^{\dagger}$ and $\tilde{g}:=\hat{U} g \hat{U}^{\dagger}$ we have the following trivial equality
\begin{equation}\label{eqqubit1}
  \C_{\alpha}(g;\O)=\C_{\alpha}(\tilde{g};\tilde{\O})\,,
\end{equation}
where  $\C_{\alpha}(\tilde{g};\tilde\O)$ denotes the complexity of $\forall\tilde{\O}\in\tilde{\mathcal{O}}$ measured by $\tilde{g}$.

In general, the gates set is not invariant under the unitary transformation, i.e.
\begin{equation}\label{ggneq}
  \tilde{g} =\hat{U} g \hat{U}^{\dagger}\neq g\,, ~~~~\text{($g$ is the gates set rather than a gate)}
\end{equation}
so we will see that the complexity of $\O$ and $\tilde{\O}$, measured by same gates set $g$, will not be the same, i.e.
\begin{equation}
\C_{\alpha}(g;\O)=\C_{\alpha}(\tilde{g};\tilde{\O})\neq\C_{\alpha}(g;\tilde{\O})
\end{equation}
This shows that the complexity for qubit system will not be invariant under $\hat{O} \rightarrow \hat{U}\hat{O}\hat{U}^\dagger$ {if we use the same gates set.}

For SU($n$) groups, we still obtain an equation similar to Eq.~\eqref{eqqubit1},
\begin{equation}\label{eqqubit2}
  \C_{\alpha}(\mathfrak{su}(n);\O)=\C_{\alpha}(\widetilde{\mathfrak{su}}(n);\tilde{\O})\,.
\end{equation}
However, unlike Eq. \eqref{ggneq} in qubit systems,  we have the following equality\footnote{The equality means the equality between two sets, i.e, $\widetilde{\mathfrak{su}}(n)$ and $\mathfrak{su}(n)$ contain the same elements.}
\begin{equation}\label{susun}
  \widetilde{\mathfrak{su}}(n):=\hat{U} \mathfrak{su}(n)\hat{U}^{\dagger}=\mathfrak{su}(n),~~\forall\x\in\text{SU($n$)}\,.
\end{equation}
Thus, we see that Eq.~\eqref{eqqubit2} implies that
\begin{equation}\label{adjointOO1}
  \C_{\alpha}(\mathfrak{su}(n);\O)=\C_{\alpha}(\mathfrak{su}(n);\tilde{\O}),~~\forall\O\in\text{SU($n$)},
\end{equation}
which means the complexity of SU($n$) group will be invariant under the adjoint transformation, $\hat{O} \rightarrow \hat{U}\hat{O}\hat{U}^\dagger$.

It is the difference between Eq.~\eqref{ggneq} and Eq.~\eqref{susun} that leads the difference between qubit systems and SU($n$) regarding the invariance under adjoint transformations. {Because  Eq.~\eqref{adjointOO1} is valid also for any infinitesimal operator, it implies Eq.~\eqref{adjointcs0}. This is another derivation of  Eq.~\eqref{adjointcs0}.}
We have presented two {arguments to support} the idea that the complexity of SU($n$) group should be invariant under adjoint transformations. In appendix~\ref{othermetric}, we will give the third and the fourth arguments to support this conclusion.

To understand the validity of the adjoint invariance of the complexity, one useful question is the following:  what will happen if we restrict our operators set to some subgroup of SU($n$)? Let $G$ to be a connected real subgroup and its Lie algebra to be $\mathfrak{g}$. In this case, we can still obtain the following equation under a general unitary transformation $\tilde{G}=\x G\x^{-1}$ and $\tilde{\mathfrak{g}}=\x\mathfrak{g}\x^{-1}$,
\begin{equation}\label{eqqubit3}
  \C_{\alpha}(\mathfrak{g};\O)=\C_{\alpha}(\widetilde{\mathfrak{g}};\tilde{\O}),~~\forall\x\in\text{SU($n$)},~\forall\O\in G\,.
\end{equation}
If $\mathfrak{g}$ is an ideal of $\mathfrak{su}(n)$, $\widetilde{\mathfrak{g}}=\mathfrak{g}$ for all $\x\in$SU($n$).  However, because the $\mathfrak{su}(n)$ is simple   it does not have other ideals except for the trivial $\{0\}$. Thus, if we restrict the operators set to any real subgroup of SU($n$), the complexity may not be invariant under a adjoint transformation. For qubit systems such as a quantum circuit, the gates set is discrete, which can form only a subgroup of SU($n$). {As SU($n$) group does not have non-trivial normal subgroup,}  the complexity for qubit systems is not invariant under the general adjoint transformation.

\subsection{Reversibility of Finsler metric from the CPT symmetry}\label{CPTinva}
In this subsection we consider the effect of the CPT symmetry of the quantum field theory on the Finsler metric, cost, and complexity.
Let us denote the CPT partner of $\Phi(\vec{x},t)$ by $\bar{\Phi}(\vec{x},t)$. i.e. $\bar{\Phi}(\vec{x},t) :=C\circ P\circ T[\Phi(\vec{x},t)]=C[\Phi(-\vec{x},-t)]$.
  By using Eq.~\eqref{dynamicPhi1}, we have
\begin{equation}\label{dynamicPhi2}
\begin{split}
  \bar{\Phi}(\vec{x},t)&= C\circ P\circ T[c(t)^\dagger\Phi(\vec{x},0)c(t)] \\
  &=c(-t)^\dagger C[\Phi(-\vec{x},0)]c(-t) \\
  &=c(-t)^\dagger \bar{\Phi}(\vec{x},0)c(-t)\,,
  \end{split}
\end{equation}
where we use the fact that $c(t)$ does not have charge and spatial variable $\vec{x}$. Thus, the evolution of the CPT parter is given by $\bar{c}(t):=c(-t)$. Given the CPT symmetry of the theory, it is natural to assume that the costs of ${c}(t)$ and $\bar{c}(t)$ are the same, i.e.,
\begin{equation}\label{arugueA1s1}
  L_\alpha[c(t)]=L_\alpha[c(-t)]\,.
\end{equation}

Similarly to the unitary symmetry case in subsection \ref{sec41}, as a way to understand the general structure of $\tilde{F}$, we  consider a special curve, the time evolution given by an arbitrary constant generator $H_\alpha$.
Because the generators of $\bar{c}(s)$ are given by $\bar{H}_\alpha=-H_\alpha$, Eq. $\eqref{arugueA1s1}$ reads, by the definition of the cost Eq. \eqref{Lalpha},

\begin{equation}\label{arugueA1s2}
  \int_0^1\F(H_\alpha)\td t=\int_0^1\F(-H_\alpha)\td t\,.
\end{equation}
which implies
\begin{equation} \label{TR11}
 \F(H_\alpha)=\F(-H_\alpha) \,.
\end{equation}

\paragraph{Path-reversal symmetry} If we combine the result of the CTP symmetry and the unitary symmetry,  Eq.~\eqref{TR11} and  Eq.~\eqref{adjinva1} respectively, one can prove the ``path-reversal symmetry'' for an arbitrary curve:
\begin{equation}\label{path-inver}
  L_\alpha[c]=L_{\alpha}[c^{-1}],  \qquad \forall c(s) \,.
\end{equation}
Note that in general $c^{-1}(s):=[c(s)]^{-1}$ is not the curve generated by $-H_\alpha(s)$ when the curve $c(s)$ is generated by $H_\alpha(s)$. For example, for the right-invariant case, we can show
\begin{equation}
\tilde{F}(H_r (c^{-1})) = \tilde{F}(- c^{-1} H_r (c) c) = \tilde{F}( H_r (c)) \,,
\end{equation}
which gives Eq. \eqref{path-inver}. Here, we used $H_r(c^{-1}) = (\td c^{-1}/\td s) c = -c^{-1}(\dot{c} c^{-1}) c =-c^{-1} H_r(c) c$ in the first equality and  Eqs. \eqref{TR11} and \eqref{adjinva1} in the second equality.
In fact, the reverse also holds, i.e. Eq. \eqref{path-inver} implies Eqs. \eqref{TR11} and \eqref{adjinva1}.
First, by considering the special case $c = e^{Hs}$ with a constant $H$, Eqs. \eqref{TR11} can be derived from  Eq. \eqref{path-inver}. Then, we end up with $L_r[c] = L_l[c]$, which implies Eq. \eqref{adjinva1} by the same logic in subsection \ref{sec41}.

Thus, we have the following equivalence between the path-reversal symmetry and the adjoint invariance with the reversibility of the Finsler metric:
\begin{equation}\label{pathCPT1}
  \forall c(s),~~L_\alpha[c]=L_{\alpha}[c^{-1}]~ \Leftrightarrow ~
  \left\{
  \begin{split}
  &\tilde{F}(H_\alpha)=\tilde{F}(\hat{U}H_\alpha\hat{U}^\dagger),~~~\forall H_\alpha, ~~ \forall\hat{U}\in\text{SU($n$)};\\
  &\tilde{F}(H_\alpha)=\tilde{F}(-H_\alpha), \qquad~ \forall H_\alpha\,.
  \end{split}
  \right.
\end{equation}
The path-reversal symmetry also can be justified by other ways, for example, based on the inverse symmetry of the relative complexity or the ``ket-world''-``bra-world'' symmetry. These two arguments are presented in appendix~\ref{othermetric} in detail. They may server other supporting evidence for Eqs. \eqref{TR11} and \eqref{adjinva1} because of the equivalence in Eq. \eqref{pathCPT1}.

\section{Complexity of SU($n$) operators}   \label{sec5}

\subsection{Finsler metric of SU($n$) operators }
From here we will drop the indexes $\alpha, r$, and $l$ based on Eq.~\eqref{bi-inv}.
We have found two constraints Eq.~\eqref{adjinva1} and Eq.~\eqref{TR11}  by considering basic physical symmetries. These constraints (plus \textbf{G3}) turn out to be strong enough to determine the Finsler metric in the operator space of any SU($n$) groups as follows
%
\begin{equation}\label{formforFs1}
\begin{split}
  F(c(s),\dot{c}(s))&= \tilde{F}(H(s)) = \lambda\text{Tr}\sqrt{H(s)H(s)^\dagger} \,, \\
  \end{split}
\end{equation}
where $H(s) = H_r(s)$ or $H_l(s)$ for the curve $c(s)$ and $\lambda$ is arbitrary constant.
(see appendix~\ref{app2} for a proof.)


The Finsler metric Eq. \eqref{formforFs1} is representation independent and to find the explicit Finsler metric tensor for the SU($n$) group we consider the fundamental representation.
An arbitrary generator $H(s)$ can be expanded as
\begin{equation}
H(s)=iH^a(s)T_a,~~~H^a(s)\in\mathbb{R}\,,
\end{equation}
where $\{T_a\}$ are basis of $\mathfrak{su}(n)$ in the fundamental representation with the following property.
\begin{equation}\label{prodTaTb}
  T_aT_b=\frac1{2n}\delta_{ab}\I+\frac12\sum_{c=1}^{n^2-1}(i{f_{ab}}^c+{d_{ab}}^c)T_c\,,
\end{equation}
where  ${f_{ab}}^c$ is the structure constants antisymmetric in all indices, while  ${d_{ab}}^c$, which is nonzero only when  $n>2$, is symmetric in all indices and  traceless.
Thus, Eq.~\eqref{formforFs1} reads
\begin{equation}\label{ftildeSUn}
  F(c,\dot{c})=\frac1{\sqrt{2n}}\Tr\sqrt{H^a(s)H^b(s)[\delta_{ab}\I+n{d_{ab}}^cT_c]} \,,
\end{equation}
with
\begin{equation}\label{FindHa1}
  H^a(s)=2\Tr[H(s)T^a]=2\Tr[\dot{c}(s)c^{-1}(s)T^a]\,,
\end{equation}
and
\begin{equation}\label{TaTa}
  T^a:=T_b\delta^{ab}\,.
\end{equation}
Without loss the generality, we have set $\lambda=1$.

For $n=2$, ${d_{ab}}^c=0$ so Eq. \eqref{ftildeSUn} is simplified to
\begin{equation}\label{ftildeSU2}
\begin{split}
  F(c,\dot{c})& =\frac12\Tr\sqrt{H^a(s)H^b(s)\delta_{ab}\I}=2\sqrt{H^a(s)H^b(s)\delta_{ab}} \\
  &= \sqrt{\Tr[\dot{c}(s)c^{-1}(s)T^a]\Tr[\dot{c}(s)c^{-1}(s)T^b]\delta_{ab}}\,,
\end{split}
\end{equation}
where Eq.~\eqref{FindHa1} is used.
%
%
%
It contains only quadratic terms of $\dot{c}$ so it gives a Riemannian geometry.  For $n>2$
\begin{equation}\label{ftildeSUn4}
\begin{split}
  F(c,\dot{c})=&\frac{2}{\sqrt{2n}} \Tr\sqrt{\Tr[H(s)T^a]\Tr[H(s)T^b][\delta_{ab}\I+n{d_{ab}}^cT_c]} \\
  =&\frac{2}{\sqrt{2n}} \Tr\sqrt{\Tr(\dot{c}c^{-1}T^a)\Tr(\dot{c}c^{-1}T^b)[\delta_{ab}\I+n{d_{ab}}^cT_c]} \,.
  \end{split}
\end{equation}
As $\I$ and $T_c$ are $n\times n$ matrixes, the line element in Eq.~\eqref{ftildeSUn4} is not the quadratic form of $\dot{c}$ so it is not Riemannian but Finsler.


\subsection{Geodesics and complexity of SU($n$) operators} \label{sec52}
Even though we have the precise Finsler metric, to compute the complexity, we still have to find a geodesic path as shown in \eqref{defcomF1}. This minimization procedure is greatly simplified thanks to {\it bi-invariance} implied by \eqref{bi-inv}. The bi-invariance means the geometry is both right and left invariant.
It has been shown that, in bi-invariant Finsler geometry, the curve $c(s)$ is a geodesic if and only if there is a constant generator $H(s) = \bar{H}$ such that~\cite{Latifi2013,Latifi2011}
\begin{equation}\label{geodesic1}
\dot{c}(s)=\bar{H}c(s) \ \  \text{or} \  \ c(s)=\exp(s\bar{H}) \,.
\end{equation}
With the condition $\hat{O}=c(1)=\exp(\bar{H})$, we can solve $\bar{H}$ formally $\bar{H}=\ln\hat{O}$. The logarithm of a unitary operator always exists but may not be unique (theorem 1.27 in Ref.~\cite{higham2008functions}).
%
%
{Because $\bar{H}$ is constant, from Eqs. \eqref{Lalpha}\,,
\begin{equation}
L[c] = \tilde{F}(\bar{H}) =  \text{Tr}\sqrt{\bar{H}\bar{H}^\dagger}  \,.
\end{equation} }
Finally, the complexity of $\hat{O}$ {in Eq.~\eqref{defcomF1}} is given by
\begin{equation}\label{compforO}
  \mathcal{C}(\hat{O}) 
  = \min\{ \text{Tr}\sqrt{\bar{H}\bar{H}^\dagger}\ | \ \forall \, \bar{H}=\ln\hat{O}\}\,,
\end{equation}
 The minimization `min' in  \eqref{defcomF1} in the sense of `geodesic' is already taken care of in  \eqref{geodesic1}.  Here `min' means the minimal value due to multi-valuedness of $\ln \hat{O}$. 

For example, let us consider the SU(2) group in the fundamental representation.  For any operator $\hat{O}\in$SU(2), there is a unit vector $\vec{n}$ and a real number $\theta$ such that,
\begin{equation}\label{relnaO}
  \hat{O} =  \exp(i \theta\vec{n}\cdot\vec{\sigma}) =\I\cos \theta+i(\vec{n}\cdot\vec{\sigma})\sin \theta  \,, 
\end{equation}
where 
$\vec{\sigma}:=(\sigma_x,\sigma_y,\sigma_z)$ stands for three Pauli matrixes.
Because $\ln\hat{O}=i \theta_m\vec{n}\cdot\vec{\sigma}$ with
\begin{equation}\label{solvesu2}
 \theta_m=\arccos[\text{Tr}(\hat{O})/2]+2m\pi \,,
\end{equation}
for $\forall m\in \mathbb{N}$, the complexity of $\hat{O}$ is given by
\begin{equation}\label{compOsu2}
  \mathcal{C}(\hat{O})= 2\arccos[\text{Tr}(\hat{O})/2] \,,
\end{equation}
where $  \bar{H}\bar{H}^{\dagger}=\theta_m^2\I $ is used.

\section{{Minimal cost principle}}  \label{sec6}

The fact that the process of complexity is generated by a constant generator allows an interesting interpretation of the Schr\"{o}dinger's equation.
\begin{figure}
  \centering
  \includegraphics[width=.5\textwidth]{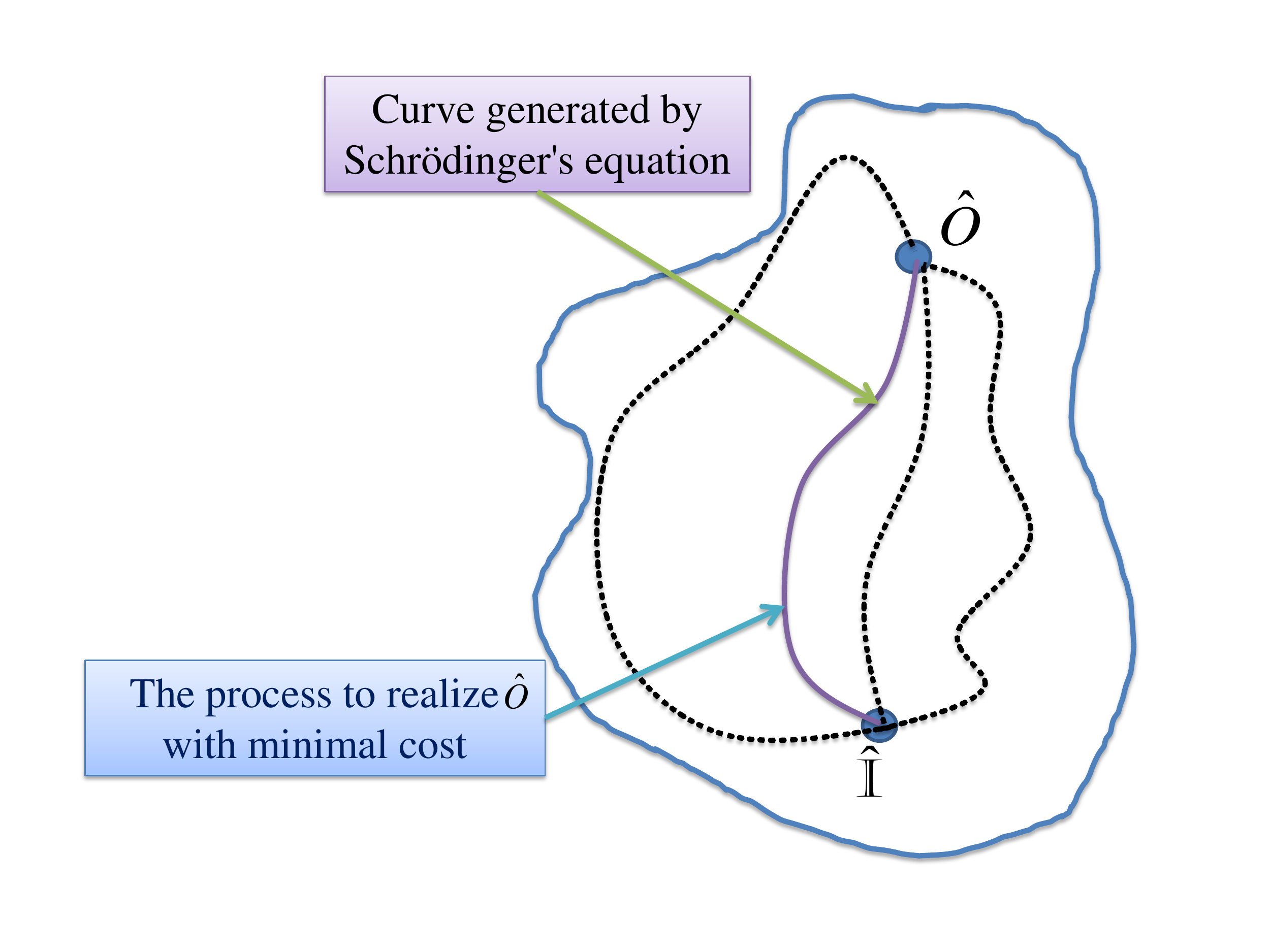}
  \caption{The schematic figure showing the curve of a time evolution operator $\hat{U}(t)$.  {To reach the same target operator at $t=t_0$, there are many different curves. The curve given by Schr\"{o}dinger's equation coincides with the curve of complexity. }}\label{expalinC1}
\end{figure}
In a quantum system with a time-independent Hamilton $\mathbb{H}$, the time evolution of the quantum state $|\psi(t)\rangle$ is given by a time evolution operator $\hat{U}(t)$, i.e. $|\psi(t)\rangle=\hat{U}(t)|\psi (0)\rangle$, where $\hat{U}(t)$ satisfies the Schr\"{o}dinger's equation,
\begin{equation}\label{schro1}
  \frac{\td}{\td t}\hat{U}(t)=-i\hbar^{-1}{\mathbb{H}}\hat{U}(t),~~~\hat{U}(0)=\I\,.
\end{equation}
Comparing with Eq.~\eqref{geodesic1}, we may say that $\hat{U}(t)$ is a geodesic generated by $-i\hbar^{-1}{\mathbb{H}}$. Thus, the time-evolution operator $\hat{U}(t)$ is along the curve of minimal (at least locally minimal) complexity.

Now let us re-consider the problem in the following way. Assume that after a short time $t=t_0$, the time-evolution operator becomes $\hat{U}(t_0)=\hat{O}$. As there are many different curves which connect the $\I$ and $\hat{O}$ (see the Figure~\ref{expalinC1}), how can we find the real curve $\hat{U}(t)$ during $t\in(0,t_0)$? One answer is that we assume the time evolution operator will obey the Schr\"{o}dinger's equation~\eqref{schro1}. Alternatively, we may replace the  Schr\"{o}dinger's equation with the following principle:

\begin{enumerate}
\item[] \textbf{Minimal cost principle:}  {For isolated systems, the time-evolution operator $\hat{U}(t)$ will be along the curve {to reach the target operator} so that the cost during this process is locally minimal, i.e. the evolution curve will make the following integral to be locally minimal:}
\begin{equation}\label{LUt1}
  L[\hat{U}(t)]=\int_0^{t_0}\text{Tr}\sqrt{\dot{\hat{U}}(t)[\dot{\hat{U}}(t)]^\dagger}\td t,~~\hat{U}(0)=\I,~~~\hat{U}(t_0)=\O\,,
\end{equation}
where we used Eq. \eqref{formforFs1}, where $ H \rightarrow    \dot{\hat{U}}(t)[\hat{U}(t)]^\dagger$.
\end{enumerate}
As a result, the time evolution operator $\hat{U}(t)$ will satisfy Eq.~\eqref{schro1}. In other words, by this principle, the Schr\"{o}dinger's equation is not the first-principle but a consequence of the {complexity principle}.\footnote{If the Hamiltonian is not static, the {complexity} curve and the curve generated by Hamiltonian will be different. 
{In this case, we may extend the system to include back-reaction of the system on its background, rendering the total Hamiltonian static.}}



\section{Comparison with the complexity for $K$-local qubit systems}\label{kqubit}  \label{sec7}

For a better understanding of the novel aspects of our work compared to previous research it is useful to compare our complexity and the complexity for $K$-qubit systems~\cite{Brown:2017jil}. In particular, our complexity is bi-invariant but the complexity geometry in Ref.~\cite{Brown:2017jil} is only right-invariant. At first glance, two theories may look different, but the difference in complexity turns out to be little and most of physical results in Ref.~\cite{Brown:2017jil} can also appear in our theory.

For $K$-qubit system,  the operators form a SU$(2^K)$ group and can be generated only by a right Hamiltonian
\begin{equation}\label{KlocalH}
  H_r(s)=\sum_IH^a_r(s)T_a\,,
\end{equation}
where $a=1,2,3,\cdots, 4^K-1$ and $T_a$ is a series of generalized Pauli operators which can span the Lie algebra $\mathfrak{su}$($2^K$).  In Ref.~\cite{Brown:2017jil},  for the SU($2^K$) group,  the following Riemannian metric was proposed
\begin{equation}\label{SU2Kg1}
  \td l^2=\td\Omega_a\mathcal{I}^{ab}\td\Omega_b\,,
\end{equation}
where
\begin{equation}\label{defOmegaI}
  \td\Omega_a:=i\Tr(\td c^\dagger(s)T_ac(s)) \,.
\end{equation}
Here, the matrix $\mathcal{I}^{ab}$ should be chosen as a block diagonal matrix with one block corresponding to the unpenalized $k$-local directions, and the other block corresponding to the directions $T_a$ containing more than $k$ single qubit operators. Note that, for given $a$ and $b$, $iT_a$ is matrix-valued vector in the representation space of $\mathfrak{su}(2^K)$ and $\mathcal{I}^{ab}$ is a real number.

The curve $c(s)$ is assumed to be generated by a right generator $H_r(s)$ such that
\begin{equation}\label{rHSUk1}
  \td c(s)=H_r(s)c(s)\td s,~~~~H_r(s)\in\mathfrak{su}(2^K)\,.
\end{equation}
%
Taking Eq.~\eqref{rHSUk1} into Eq.~\eqref{defOmegaI}, we have
\begin{equation}\label{defOmegaI1}
\begin{split}
  \td\Omega_a&=\Tr[c^\dagger(s)H_r(s)T_ac(s)]\td s\\
  &=\Tr[H_r(s)T_a]\td s \,,
  \end{split}
\end{equation}
%
so Eq. \eqref{SU2Kg1} becomes
\begin{equation}\label{SU2Kg3}
  \td l^2=\Tr[H_r(s)T_a]\Tr[H_r(s)T_a]\mathcal{I}^{ab}\td s^2\,,
\end{equation}
which yields the following Finsler (or just Riemannian) metric
\begin{equation}\label{RFmetric}
F(c,\dot{c})= \sqrt{\Tr[H_r(s)T_a]\Tr[H_r(s)T_b]\mathcal{I}^{ab}}\,.
\end{equation}
It has two differences compared with our result \eqref{ftildeSUn4}. One is that the ``$\Tr$'' is inside the square root. The other is that there is a matrix structure $\mathcal{I}^{ab}$ which is not determined by the Lie algebra uniquely.   In the following, we make comparisons between two complexities based on two different Finsler metrics.

\begin{enumerate}
\item[(1)]In our paper, the only basic assumptions are \textbf{G1}-\textbf{G4}. All conclusions such as Finsler geometry and the Finsler metric Eq.~\eqref{formforFs1} are the results of these four assumptions and fundamental  symmetries of QFTs. In Ref.~\cite{Brown:2017jil}, the Riemannian geometry and metric \eqref{SU2Kg3} in $K$-qubit system were proposed directly as the basic assumptions.
\item[(2)]The complexity given by Eq.~\eqref{RFmetric} satisfies our axioms \textbf{G1} and \textbf{G2} but breaks \textbf{G3}. It can be shown by considering a simple case, $\mathcal{I}^{ab}=\delta^{ab}$, which corresponds to bi-invariant case without any penalty (isotropy).   {In this case, the complexity of the operator $\hat{O} = {\text{exp}}(H)$ is given by $F(H)$ because the geodesic is generated by a constant generator (due to bi-invariance)  i.e.
\begin{equation} \label{xyz1}
C(\hat{O}) = F(H) \,.
\end{equation}
}
By using Eqs. \eqref{xyz1} and \eqref{RFmetric} we have
\begin{equation}\label{klocalG3}
  \C(\O_1\oplus\O_2)=\sqrt{\C^2(\O_1\oplus\I_2)+\C^2(\I_1\oplus\O_2)}=\sqrt{\C^2(\O_1)+\C^2(\O_2)} \,,
\end{equation}
and in more general cases,
\begin{equation}\label{klocalG3b}
  \C(\bigoplus_i\O_i)=\sqrt{\sum_ip_i[\C(\O_i)]^2}\neq\sum_i\C(\O_i)\,,
\end{equation}
where $p_i$ is a weighting factor if $\mathcal{I}^{ab}\neq\delta^{ab}$. {This means that the total complexity of \textit{parallel} operations is not the sum of the the complexity of the individual operations, so breaks \textbf{G3}. We want to stress again that \textbf{G3} is a very natural requirement that has not been considered in previous research.}
%
%
\item[(3)] For the same curve in SU($n$) group, the tangent vector at a point is unique but the generator is not. It can be a left or right generator.  We admit of two ways (left generator or right generator)  but Ref.~\cite{Brown:2017jil} considers only one way (right generator).
As there is no  reason to assume that physics favors ``left world'' or ``right world'', a simple and natural possibility is that two generators yield the same complexity. It is the case that is realized in our framework unlike the  complexities in~\cite{Brown:2017jil} and Neilsen's original works~\cite{Nielsen1133,Nielsen:2006:GAQ:2011686.2011688,Dowling:2008:GQC:2016985.2016986} where the left and right generator will give different complexities.

One may argue that Eq.~\eqref{RFmetric} could be valid only for the right generator and, for the left generator, there might be another left-invariant metric which has different penalty $\mathcal{I}'^{ab}$ and could give out the same curve length. In our upcoming work Ref.~\cite{Yang:2018tpo}, we will show this is possible only if the geometry is bi-invariant.  This gives us  another argument that the complexity for SU($n$) group should be bi-invariant. 

\item[(4)] When Ref.~\cite{Brown:2017jil} discusses some particular physical situations such as ``particle on complexity geometry'', ``complexity equals to action'' and ``the complexity growth'', the authors restricted the generators in a ``$k$-local'' subspace  $\mathfrak{g}_k$ and assumed $\mathcal{I}_{ab}|_{\mathfrak{g}_k}=\delta_{ab}$ (see the Sec.  IV.C and V in Ref.~\cite{Brown:2017jil} for detailed explanations). As a  result, the geodesics in the sub-manifold generated by $\mathfrak{g}_k$ are also given by constant generators, which are the same as our bi-invariant Finsler geometry. The lengths of such geodesics in Ref.~\cite{Brown:2017jil} and in our paper are only different by multiplicative factors, which implies that all the results given by ``$k$-local'' subspace can also appear similarly in our bi-invariant Finsler geometry.

    {Moreover, in order to obtain the complexity geometry as was proposed in Ref.~\cite{Brown:2017jil},  we can choose some two-dimensional sub-manifold in SU($n$) geometry. As described in our upcoming paper~\cite{Yang:2018tpo}, by using the Gauss-Codazzi equation, we can show that the sub-manifold can have negative induced sectional curvature somewhere despite the SU($n$) geometry is positively curved. So it satisfies the same properties as shown in Ref.~\cite{Brown:2017jil}, where the sectional curvature is made negative near the identity by choosing an appropriate penalty factor.}

\end{enumerate}

\section{Discussion and outlook}   \label{sec8}
In this paper we proposed four basic axioms for the complexity of operators:  nonnegativity (\textbf{G1}), series decomposition rule (triangle inequality)  (\textbf{G2}), parallel decomposition rule (\textbf{G3}) and smoothness (\textbf{G4}). Combining these four axioms  and basic symmetries in QFT, we have obtained the complexity of the SU($n$) operator without ambiguity: Eq.\eqref{compforO}. In our derivation the bi-invariance of the Finsler structure plays an important role and this bi-invariance is a natural implication of the symmetry in QFTs rather than an artificial assumption.
Our logical flows are shown in the Fig.~\ref{logic1}.

%

We argue the importance of the bi-invariance in four ways based on: i)  the unitary-invariance of QFTs (section \ref{sec41}); ii) the nature of continuous operators rather than discrete ones (section \ref{sec42});  iii) inverse-invariance of the relative complexity (appendix \ref{othermetric1}); and iv)  the ``ket-world'' - ``bra-world'' equivalence (appendix \ref{othermetric2}).
The bi-invariance here is different from the only right-invariance for qubit systems~\cite{Nielsen1133,Nielsen:2006:GAQ:2011686.2011688,Dowling:2008:GQC:2016985.2016986,Brown:2017jil}.  We clarify the differences and similarities of our proposal (bi-invariance) from previous researches (only right-invariance) in subsection \ref{sec42} and section \ref{sec7}.   It can be shown that most of results in only right-invariant complexity geometry proposed by Ref.~\cite{Brown:2017jil} can also appear in our framework. We want to  emphasis that the complexity cannot be a well-defined physical observable in general finite dimensional systems if the complexity geometry is not bi-invariant.

Thanks to the bi-invariance of the Finsler metric the process of minimal cost (complexity) is generated by a constant generator. This observation leads us to make a novel interpretation for the Schr\"{o}dinger's equation: the quantum state evolves by the process of minimizing ``computational cost,'' which we call `` minimal cost principle.''

As an application of the complexity of the SU($n$) operator, 
the complexity between two states described by density matrices $\rho_1$ and $\rho_2$ may be defined naturally as
\begin{equation}\label{defiCstates}
  \C(\rho_1,\rho_2):=\min\{\C(\hat{O})\, |\, \rho_2=\hat{O}\rho_1\hat{O}^{\dagger},~\forall\hat{O}\in\text{SU}(n)\}\,.
\end{equation}
%
%
In our forthcoming paper~\cite{WIP} our proposal turns out to be general enough to include and unify other recent developments for the complexity in QFT:  cMERA tensor network~\cite{PhysRevLett.115.180405,PhysRevLett.115.200401,Miyaji:2015fia,Molina-Vilaplana:2018sfn}, Fubini-Study metric~\cite{Chapman:2017rqy}
and path-integral method~\cite{Caputa:2017urj,Caputa:2017yrh}. Furthermore, it can be shown that our proposal also correctly reproduces the holographic complexity for thermofield double state (TFD).

In a more general context, geometrizing the complexity in continuous operators sets amounts to giving positive homogeneous norms in some Lie algebras. Our paper deals with only SU($n$) group so we gives the norm for Lie algebra $\mathfrak{su}(n)$. For more general Lie algebra $\mathfrak{g}$, though we cannot determine the norm uniquely, it is natural that such a norm is determined only by the properties of $\mathfrak{g}$, for example the structure constants, without any other extra information. As in general relativity where the spacetime metric is determined by matter distribution through Einstein's equations, can we find any  physical equation to determine this norm?



\acknowledgments
The work of K.-Y. Kim and C. Niu was supported by Basic Science Research Program through the National Research Foundation of Korea(NRF) funded by the Ministry of Science, ICT $\&$ Future Planning(NRF- 2017R1A2B4004810) and GIST Research Institute(GRI) grant funded by the GIST in 2019. C.Y. Zhang is supported by National Postdoctoral Program for Innovative Talents BX201600005.


\appendix

\section{Brief introduction to the Finsler manifolds}\label{introF}
This appendix introduces some basic concepts in the Finsler geometry. It is not meant to be a complete or rigorous introduction. The readers can find more details in textbooks such as Refs.~\cite{038798948X,9810245319,xiaohuan2006an} and a physics friendly introduction in Ref.~\cite{asanov1985finsler}.

\subsection{Fundamentals}

Suppose that $M$ is an $n$-dimensional smooth manifold and $TM$ is its tangent bundle. Each element of $TM$ is given by a pair $(x^i,v^i)$, where $x^i\in M$ and $v^i\in T_xM$. In this appendix, $x,y$ will be used to stand for the points in $M$ and $u,v$ will be used to stand for the tangent vectors at some points in $M$. For convenience, sometimes  their indices will be dropped if there is no ambiguity.

A Finsler metric of $M$ is a function $F: TM\mapsto[0,\infty)$ such that:
\begin{enumerate}
\item[(1)] $F$ is smooth on $TM\setminus\{\cdot, 0\}$;
\item[(2)] $F(x,\lambda v)=\lambda F(x,v)$ for arbitrary $\lambda>0$;
\item[(3)] The fundamental tensor (metric)
\begin{equation}\label{defgij}
  g_{ij}(x,v):=\frac12\frac{\partial^2F^2(x,v)}{\partial v^i\partial v^j}
\end{equation}
is positive definite when $v\neq0$. 
\end{enumerate}

The manifold $M$ with a Finsler metric $F$ is called a \textit{Finsler manifold}. 
The requirement (3) can be relaxed in physics. If $g_{ij}(x,v)$ has negative eigenvalues but no zero eigenvalue, then $(M,F)$ is call a \textit{pseudo-Finsler manifold}; if $g_{ij}(x,v)$ has only one negative eigenvalue and has no zero eigenvalue, then $(M,F)$ is called \textit{pseudo-Finsler spacetime}. In this paper, we only consider the case that $g_{ij}(x,v)$ is positive definite.

We want to emphasize that the metric tensor $g_{ij}(x,v)$ is defined in the tangent bundle $TM$ rather than the manifold $M$, which is the essential difference between a general Finsler manifold and Riemannian manifold. The relation between the Finsler metric $F$ and the metric tensor $g_{ij}$ reads
\begin{equation}\label{relFg}
  F^2(x,v)=g_{ij}(x,v)v^iv^j\,.
\end{equation}
For a curve $x(t)=c(t)$ in $M$, its line element is given by
\begin{equation}\label{curveL1}
  \td s:=\sqrt{g_{ij}(c,\dot{c})\dot{c}^i\dot{c}^j}\td t=F(c,\dot{c})\td t\,,
\end{equation}
and the curve length for $t\in(0,1)$ is 
\begin{equation}\label{curveL1}
  L[c(t)]:=\int_0^1F(c,\dot{c})\td t=\sqrt{g_{ij}(c,\dot{c})\dot{c}^i\dot{c}^j}\td t\,.
\end{equation}

In a Riemannian case, the curves $x(t)=c(t)$ and $x(t)=c(1-t)$ with $t\in(0,1)$ have the same length. However, this is no longer true for general Finsler geometries because
\begin{equation}\label{curveL1}
  L[c(1-t)]:=\int_0^1\sqrt{g_{ij}(c,-\dot{c})\dot{c}^i\dot{c}^j}\td t\neq\int_0^1\sqrt{g_{ij}(c,\dot{c})\dot{c}^i\dot{c}^j}\td t=L[c(t)]\,.
\end{equation}
%
%
If the metric $g_{ij}(x,v)$ is independent of $v$, then a Finsler manifold becomes a Riemannian manifold. To describe how much different a  Finsler manifold is from a Riemannian manifold, we can introduce the Cartan's tensor,\footnote{In some textbook, there is an overall factor $F$ in the definition so that the Cartan tensor is scale invariant under the transformation $v\rightarrow\lambda v$.}
\begin{equation}\label{Cartan1}
  A_{ijk}(x,v) := \frac12\frac{\partial }{\partial v^k}g_{ij}(x,v)=\frac14\frac{\partial^3F^2(x,v)}{\partial v^i\partial v^j\partial v^k}\,,
\end{equation}
which is a fully symmetric covariant tensor.

Thus, a Finsler manifold is a Riemannian manifold if and only if its Cartan tensor is zero. The Cartan form ($\eta_k$) is defined by the contraction of Cartan tensor and metric:
\begin{equation}\label{defA2eta}
  \eta_k(x,v):=g^{ij}(x,v)A_{ijk}(x,v)\,,
\end{equation}
where $g^{ij}(x,v)$ is the inverse of $g_{ij}(x,v)$. The Cartan tensor and Cartan form have three important properties:
\begin{equation}\label{Cartan1}
  \eta_k(x,v)=\frac{\partial}{\partial v^k}\ln\sqrt{\det(g_{ij})},~~~u=v\Rightarrow A_{ijk}(x,v)u^j=0,~~~\eta_k=0\Leftrightarrow A_{ijk}=0\,.
\end{equation}

\subsubsection*{Example}
In order to give readers a ``feeling'' about the Finsler geometry, let us show a simple example. Assume $a_{ij}(x)$ to be a positive definite tensor and a 1-form $b_i(x)$ satisfying $b_i(x)b_j(x)a^{ij}(x)<1$.\footnote{This condition is necessary to have a positive definite metric tensor.} With $\alpha(x,v)=\sqrt{a_{ij}(x)v^iv^j}$ and $ \beta(x,v)=b_i(x)v^j$, 
\begin{equation}\label{Rander}
  F(x,v)=\alpha(x,v)+\beta(x,v)
\end{equation}
is a Finsler metric. If $b_i(x)=0$, this is just a Riemannian geometry. By the definition~\eqref{defgij}, the metric tensor reads
\begin{equation}\label{metricRander}
  g_{ij}(x,v)=\frac{F}{\alpha}\left(a_{ij}-\frac{v_i}{\alpha}\frac{v_j}{\alpha}\right)+\left(\frac{v_i}{\alpha}+b_i\right)\left(\frac{v_j}{\alpha}+b_j\right)\,,
\end{equation}
where $v_i:=a_{ij}v^j$. One can check $g_{ij}(x,v)v^iv^j=F(x,v)^2$ and the Cartan form is
\begin{equation}\label{Cartan2}
  \eta_k(x,v)=\frac{n+1}{2F}\left(b_k-\frac{\beta}{\alpha^2}v_k\right)\,.
\end{equation}
It can be shown that $\forall v, \eta_k(x,v)=0\Leftrightarrow b_k(x)=0\Leftrightarrow g_{ij}(x,v)=a_{ij}(x)$. The Finsler geometry \eqref{Rander} is called Randers geometry, which was first proposed by physicist Gunnar Randers regarding the uni-direction of time in  general relativity~\cite{PhysRev.59.195}.

In Finsler geometry, we can also define some geometric quantities such as connection and curvature. As this paper will not consider such quantities, we will not introduce them. The readers can refer to some textbooks, e.g., Refs.~\cite{038798948X,9810245319,xiaohuan2006an}.
\subsection{Invariant Finsler geometries for a Lie group}
Now let us consider the case the manifold $M=G$ is a Lie group with the identity $\I$. The Lie algebra is denoted by $\mathfrak{g}$, which is also the tangent space of identity $\I$, i.e., $\mathfrak{g}=T_{\I}G$. Because this manifold has group structure, i.e., the product rule, we can pick out a group-relevant special Finsler metric.

In a Lie group $G$, we can define a left translation $\mathcal{L}_x: G\mapsto G$ such that $\forall y\in G, \mathcal{L}_x(y)=xy$. As the left translation is the isomorphism of the group $G$, it is natural to expect that it is also the isometry of $G$. We call a Finsler manifold $(G,F)$ is left-invariant if the left translation is an isometry, i.e.,
\begin{equation}\label{defleft-inv}
  \forall x\in G,~~~F(y,v)=F(\mathcal{L}_x(y),\mathcal{L}_{x*}(v)) \,,
\end{equation}
where $\mathcal{L}_{x*}$ is the push-forward map induced by $\mathcal{L}_x$ and  $\mathcal{L}_{x*}(v)=xv$. Thus, the left-invariant Finsler metric will satisfy
\begin{equation}\label{defleft-inv2}
  \forall x\in G,~~~F(y,v)=F(xy,xv)
\end{equation}
Particularly, by taking $x=y^{-1}$ we have
\begin{equation}\label{defleft-inv3}
  \forall y\in G,~~~F(y,v)=F(\I,y^{-1}v) \,.
\end{equation}
Similarity, we can define the right translation and right-invariant Finsler geometry. The right-invariant Finsler geometry satisfies
\begin{equation}\label{defright-inv3}
  \forall y\in G,~~~F(y,v)=F(\I,v y^{-1}) \,.
\end{equation}
Note that both left-invariant and right-invariant Finsler geometry are determined by the group product rule and the Finsler metric at the identity. A Finsler manifold is called bi-invariant if the Finsler geometry is both left and right invariant.

 The Finsler metric at the identity can be used as a norm at the Lie algebra $\mathfrak{g}$. For a given Finsler metric at the identity, we have two natural ways to obtain the Finsler geometry according to Eqs.~\eqref{defleft-inv3} and \eqref{defright-inv3}. The left and right invariant Finsler metric $F_l$ and $F_r$ can be determined uniquely
\begin{equation}\label{defFrFr}
  F_l(x,v)=\F(x^{-1}v)\,, \qquad F_r(x,v)=\F(vx^{-1})\,,
\end{equation}
where  $\F$ denotes a norm for Lie algebra $\mathfrak{g}$. 
Thus, if we know how to define the norm $\F$ for a Lie algebra $\mathfrak{g}$, we can obtain two natural Finsler geometries for the Lie group $G$. Such two Finsler geometries are the same only when the norm $\F$ satisfies
\begin{equation}\label{defbiinvs1}
  \forall v\in\mathfrak{g}, \forall x\in G,~~~\F(v)=\F(xvx^{-1})\,.
\end{equation}
In this case, we can find $F_l=F_r$ and so the Finsler geometry is bi-invariant. 

If we know the norm $\F$ for Lie algebra $\mathfrak{g}$, then for arbitrary curve $c(s)$ with $s\in(0,1)$, we have two ways to compute its length
\begin{equation}\label{curveLc1}
  L_\alpha[c]:=\int_0^1F_\alpha(c(s),\dot{c}(s))\td s\,.
\end{equation}
If the curve is generated by the left generator $H_l(s)$ and right generator $H_r(s)$ defined as
\begin{equation}\label{defleftH}
c(s)H_l(s)=\dot{c}(s),~~~H_r(s)c(s)=\dot{c}(s) \,, 
\end{equation}
we see that Eq.~\eqref{curveLc1} reads
\begin{equation}\label{curveLc2}
  L_\alpha[c]=\int_0^1\F(H_\alpha(s))\td s\,.
\end{equation}
In general, we have two different lengths for the same curve if we obtain the norm $\F$ of the Lie algebra $\mathfrak{g}$. Such two lengths are the same if $\F$ satisfies Eq.~\eqref{defbiinvs1}, i.e., the Finsler geometry is bi-invariant.

\section{Proof for the emergence of the Finsler metrics}\label{app1}
In this appendix, we first show three properties \textbf{F1-F3} of the Minkowski norm, $\tilde{F}$,  implied by the general axioms \textbf{G1}, \textbf{G2} and the assumption \textbf{G4}. \textbf{F1-F3} readily show that the Finsler geometry  $F_{\alpha}(c,\dot{c})$ emerges. {As we have clarified below Eq. \eqref{compdel}, for an infinitesimal operator $\delta\O=\exp(H\varepsilon)$ with $|\varepsilon|\ll1$, the indexes $\alpha=\{r,l\}$ are not necessary, so we omit it not to clutter notations. } \\

\noindent\textbf{F1}: $\forall H$,  $\tilde{F}(H)\geq0$ and $\tilde{F}(H)=0\Leftrightarrow H=0$.\\
\emph{Proof: }\\  
\textcircled{1} If $H=0$, $\exp(Hs)=\I$ so $\C(\exp(Hs))=0$ for $\forall s\in[0,1]$ by \textbf{G1}.   Thus, $\F(H)=\lim_{s\rightarrow0^+}\C(\exp(Hs))/s=0$ by \eqref{compdel}. $\square$  \\
\textcircled{2} If $H\neq0$, $\exists\lambda>0$ such that $\exp(H\lambda)\neq\I$ so by \textbf{G1}%
\begin{equation*}
  0<\C(\exp(H\lambda))=\C((\exp(H\lambda/N))^N)\,,
\end{equation*}
where $N\in\mathbb{N}^+$. By \textbf{G2} we have the following inequality
\begin{equation*}
  \C[(\exp(H\lambda/N))^N]\leq N\C[\exp(H\lambda/N)]\,.
\end{equation*}
Thus,
\begin{equation}\label{ineqforNC1}
 0 < \lim_{N\rightarrow\infty}\C[(\exp(H\lambda/N))^N] \le  \lim_{N\rightarrow\infty}N\C[\exp(H\lambda/N)] \,.
\end{equation}
With $\varepsilon:=\lambda/N$,  Eq.~\eqref{ineqforNC1} means
\begin{equation}\label{ineqforNC2}
  0<\lim_{\varepsilon\rightarrow0^+}\lambda\frac{\C[\exp(H\varepsilon)]}{\varepsilon}=\lambda\F(H)\,,
\end{equation}
where $\F$ is defined in \eqref{compdel}. Thus, $\F(H)>0$.
$\square$ \vspace{-0.2cm}\\

\noindent \textbf{F2}: $\forall\lambda\in\mathbb{R}^+$, $\tilde{F}(\lambda H)=\lambda\tilde{F}(H)$.\\
\emph{Proof: } \\
By \eqref{compdel}, for an arbitrary generator $H$ and infinitesimal parameter $\varepsilon>0$
\begin{equation}\label{proofp1a}
  \mathcal{C}(\exp(H\varepsilon))=\tilde{F}(H)\varepsilon+\mathcal{O}(\varepsilon^2)\,.
\end{equation}
For an arbitrary $\lambda>0$,
\begin{equation}
\mathcal{C}[\exp(H\cdot\lambda\varepsilon)]=\mathcal{C}[\exp(\lambda H\cdot\varepsilon)] \,,
\end{equation}
which implies
%
%
%
%
$\tilde{F}(\lambda H)=\lambda\tilde{F}(H)$ by Eq.~\eqref{proofp1a}.
$\square$ \vspace{-0.2cm} \\

\noindent\textbf{F3}:  $\forall H_1\neq0$ and $\forall H_2\neq0$, $\tilde{F}(H_1)+\tilde{F}(H_2)\geq\tilde{F}(H_1+H_2)$\\
\emph{Proof: } \\
By \textbf{G2}, for arbitrary generators $H_1$ and $H_2$  
\begin{equation*}\label{proofp3a}
  \mathcal{C}(\exp(H_1\varepsilon))+\mathcal{C}(\exp(H_2\varepsilon))\geq\mathcal{C}(\exp(H_1\varepsilon)\exp(H_2\varepsilon)) \,.
\end{equation*}
It yields, up to order $\mathcal{O}(\varepsilon^2)$,
\begin{equation}\label{proofp3b}
  \mathcal{C}(\exp(H_1\varepsilon))+\mathcal{C}(\exp(H_2\varepsilon))\geq\mathcal{C}[\exp((H_1+H_2)\varepsilon)]\,,
\end{equation}
%
%
which implies  $\tilde{F}(H_1)+\tilde{F}(H_2)\geq\tilde{F}(H_1+H_2)$ by Eq.~\eqref{proofp1a}. $\square$ \vspace{-0.2cm} \\ 

By the relation between $F_{\alpha}$ and $\tilde{F}$ in Eq.~\eqref{rlinvar}, we can also prove that $F_{\alpha}$ satisfies the following properties for $\forall \hat{O}\in$SU($n$) and two arbitrary tangent vectors $V, W$ at $\hat{O}$:\\
\textbf{F1'}: $F_{\alpha}(\hat{O},V)\geq0$ and $F_{\alpha}(\hat{O},V)=0\Leftrightarrow V=0$ \,. \\
\textbf{F2'}: $\forall\lambda\in\mathbb{R}^+$, we have $F_{\alpha}(\hat{O},\lambda V)=\lambda F_{\alpha}(\hat{O},V)$ \,. \\
\textbf{F3'}: $F_{\alpha}(\hat{O},V)+F_{\alpha}(\hat{O},W)\geq F_{\alpha}(\hat{O},V+W)$.\\
These imply that $F_\alpha(c,\dot{c})$ is Finsler metric.





\section{General Finsler metric for SU($n$): proof of  Eq. \eqref{formforFs1}}\label{app2}


\emph{Proof: }  We prove \eqref{formforFs1} by three steps.\\
(1)  Let us first show that $\tilde{F}(H)$ is only the function of eigenvalues of $H$ and independent of the permutations of the eigenvalues. Notice that $H$ always can be diagonalized under the transformation by an SU($n$) operator $\hat{U}$:
\begin{equation}\label{Hgamma}
\hat{U}H\hat{U}^{-1} =   \text{diag}(\gamma_1,\gamma_2,\cdots,\gamma_n) \,,
\end{equation}
where $\gamma_i$ are the eigenvalues of $H$. Thus
\begin{equation}\label{Fgamma1}
  \tilde{F}(H)=\tilde{F}(\hat{U}H\hat{U}^{-1})=\tilde{F}(\text{diag}(\gamma_1,\gamma_2,\cdots,\gamma_n))\,.
\end{equation}
where the first equality comes from the adjoint invariance \eqref{bi-inv}. To see the independence of the permutations of the eigenvalues, we consider a SU($n$) operator $\hat{U}'$ yielding
\begin{equation}\label{Hgamma2}
  \hat{U}'\text{diag}(\gamma_1,\gamma_2,\cdots,\gamma_n)\hat{U}'^{-1}=\text{diag}(\gamma_2,\gamma_1,\gamma_3,\cdots,\gamma_n)\,.
\end{equation}
%
%
where $\hat{U}'_{11}=\hat{U}'_{22}=0, \hat{U}'_{12}=-\hat{U}'_{21}=1$ and $\hat{U}'_{ij}=\delta_{ij}$ for $i,j>2$. Again by \eqref{bi-inv} $\tilde{F}$ is invariant under the permutation $\gamma_1\leftrightarrow\gamma_2$. Similarity,  $\tilde{F}[\text{diag}(\gamma_1,\gamma_2,\cdots,\gamma_n)]$ is invariant under any kind of permutation on $\{\gamma_1,\gamma_2,\cdots,\gamma_n\}$.

~\\
(2) Next, let us prove that if $H=H_1\oplus H_2$ then $\tilde{F}(H)=\tilde{F}_1(H_1)+\tilde{F}_2(H_2)$, where $\tilde{F}_1(H_1):=\tilde{F}(H_1\oplus\mathbf{0}_{n-k})$, $\tilde{F}_2(H_2):=\tilde{F}(\mathbf{0}_{k}\oplus H_2)$, $k=$Dim$(H_1)$ and $\mathbf{0}_i$ stands for the $i$-dimensional zero matrix.

Assume $\hat{O}_i$  ($i$=1,2) to be the representations of $\hat{x}_i\in\mathcal{N}_i$ and one matrix representation of $(\hat{x}_1,\hat{x}_2)\in\mathcal{N}_1\times\mathcal{N}_2$  can be given by $\hat{O}=\hat{O}_1\oplus\hat{O}_2$. Then the axiom \textbf{G3} implies that the complexity should satisfy
\begin{equation}\label{Cforx1x2}
  \mathcal{C}(\hat{O}_1\oplus\hat{O}_2)=\mathcal{C}_1(\hat{O}_1)+\mathcal{C}_2(\hat{O}_2)\,.
\end{equation}
Here {$\mathcal{C}_1(\hat{O}_1):=\mathcal{C}(\hat{O}_1\oplus\I_2)$ and $\mathcal{C}_1(\hat{O}_1):=\mathcal{C}(\I_1\oplus\hat{O}_2)$}.
To be specific, let us consider two 1-parameter subgroups $\mathcal{N}_1=\exp(sH_1)$ and $\mathcal{N}_2=\exp(sH_2)$ with $s\in\mathbb{R}$. Then $\mathcal{N}:=\mathcal{N}_1\times\mathcal{N}_2$ is the $n$-dimensional subgroup of $\mathcal{O}$.
As $H=H_1\oplus H_2$, $\exp(\epsilon H)=\exp(\epsilon H_1)\oplus\exp(\epsilon H_2)$.  From  Eq.~\eqref{Cforx1x2}, we have
\begin{equation}\label{reqCH1H2}
  \mathcal{C}[\exp(\epsilon H)]=\mathcal{C}_1[\exp(\epsilon H_1)]+\mathcal{C}_2[\exp(\epsilon H_2)] \,,
\end{equation}
which, in the limit $\epsilon\rightarrow0$, reduces to
%
\begin{equation}\label{reqCH1H22}
  \tilde{F}(H)=\tilde{F}(H_1\oplus H_2)=\tilde{F}_1(H_1)+\tilde{F}_2(H_2)\,.
\end{equation}
In general, $H=\oplus_{j}H_j$ so Eq.~\eqref{reqCH1H22} can be generalized as $\tilde{F}=\sum_j\tilde{F}_j(H_j)$.

~\\
(3) Finally, we combine the results in previous two steps. Noting the fact that $\text{diag}(\gamma_1,\gamma_2,\cdots,\gamma_n)=\bigoplus_{j=1}^n\gamma_j$, we have {
\begin{equation}\label{FFgamma}
  \tilde{F}(H) 
  =\tilde{F}\left(\bigoplus_{j=1}^n\gamma_j\right)=\sum_{j=1}^n\tilde{F}_j(\gamma_j)\,,
\end{equation}
where
\begin{equation}\label{deffis}
  \tilde{F}_j(\gamma_j): = 
  \tilde{F}(\text{diag}(\underbrace{0,0,\cdots,0}_{j-1},\gamma_i,\underbrace{0,\cdots,0}_{n-j}))\,.
\end{equation}
%
%
%
Because $\tilde{F}$ is invariant under the permutations of diagonal elements as shown in \eqref{Hgamma2}, $\tilde{F}_j(\gamma_j)$ is independent of the position of $\gamma_j$, which means there is a non-negative function $f$ such that $\tilde{F}_j(\gamma_j)=f(\gamma_j)$}. Thus
\begin{equation}\label{FFgamma2}
  \tilde{F}(H)=\sum_{j=1}^nf(\gamma_j) =\sum_{j=1}^nf(i\text{Im}\gamma_j) = \sum_{j=1}^nf(i |\gamma_j|) \,,
\end{equation}
where  we used the fact that the eigenvalues of $H$ are all pure imaginary: $\gamma_j=i\text{Im}\gamma_j$ and
$f(i\text{Im}\gamma_j)=f(-i\text{Im}\gamma_j)$ from   \eqref{const22}.
%
%
{Because the Finsler metric satisfies the homogeneity $\tilde{F}(\lambda H)=\lambda\tilde{F}(H)$  for $\lambda>0$, one can find that $f(\gamma) =|\gamma|f(i)$ for arbitrary pure imaginary number $\gamma$,\footnote{{This can be obtained, for example, by setting $H=\text{diag}(\gamma,-\gamma,0,0,\cdots,0)$ in Eq.~\eqref{FFgamma2}. }} which yields }
\begin{equation}\label{FFgamma2b}
  F(c,\dot{c})=\tilde{F}(H)=f(i)\sum_{j=1}^n|\gamma_j|=f(i)\text{Tr}\left(\sqrt{HH^\dagger}\right)\,,
\end{equation}
where, without loss of generality, we may set the overall constant $f(i)=\lambda$.
It proves \eqref{formforFs1}. $\square$

\section{{Other arguments for path-reversal symmetry}}\label{othermetric}
\subsection{{Inverse-invariance} of relative complexity}\label{othermetric1}
 Let us denote the set of all admitted gates $g_i$  by $g$ and  the set of all $g_i^{-1}$ by $g^{-1}$, i.e.,
\begin{equation}\label{definvg}
 g := \{g_i| \forall g_i\in g\}\,, \qquad  g^{-1}:=\{g_i^{-1}| \forall g_i\in g\}\,,
\end{equation}
In discrete qubit systems, if $g_i$ is one admitted gate then $g_i^{-1}$ is also one admitted gate because quantum circuits are invertible, so $g^{-1}=g$.
Let us investigate what  this property implies for the circuit complexity.

For the unitary operators pair $(\hat{U},\hat{V})$ and the fundamental gates set $g$,
the ``relative complexity'' $d_\alpha(g;\hat{U},\hat{V})$  is defined by the minimal number of required gates to transform from $\hat{V}$ to $\hat{U}$ under the gates set $g$. The index $\alpha$ is $l$ or $r$, which is explained in section \ref{sec3}.
The relative complexities can be written in terms of the complexity as
\begin{equation}\label{defrelC1}
  d_r(g;\hat{U},\hat{V})=\C_r(g;\hat{U}\hat{V}^{-1}), \qquad d_l(g;\hat{U},\hat{V})=\C_l(g;\hat{V}^{-1}\hat{U})\,.
\end{equation}
where $\C_\alpha(g;\hat{U})$ is the complexity of $\hat{U}$ under the gates set $g$,
As $g_i$ and $g_i^{-1}$ are both the elements in $g$, $\C_\alpha(g;\O)=\C_\alpha(g;\O^{-1})$ so it follows that $d_\alpha(g;\hat{U},\hat{V})=d_\alpha(g;\hat{V},\hat{U})$.

Now let us consider the relative complexities of $(\hat{U},\hat{V})$ and $(\hat{U}^{-1},\hat{V}^{-1})$.
If $\{g_1,g_2,\cdots,g_n\}$ are some gates transforming $\hat{V}$ to $\hat{U}$ by the right-invariant way $(\alpha=r)$, i.e.,
\begin{equation}\label{dUVg1}
  \hat{U}=g_ng_{n-1}\cdots g_2g_1\hat{V} \,,
\end{equation}
we have
\begin{equation}\label{dUVg2}
  \hat{U}^{-1}=\hat{V}^{-1}g_1^{-1}g_2^{-1}\cdots g_{n-1}^{-1}g_n^{-1}=\tilde{g}_1\tilde{g}_2\cdots \tilde{g}_{n-1}\tilde{g}_n\hat{V}^{-1}\,,
\end{equation}
where $\tilde{g}_i :=\hat{V}^{-1}g_i^{-1}\hat{V}$.
By defining a new gates set
$$\tilde{g}:=\{\hat{V}^{-1}g_i^{-1}\hat{V}|\forall g_i\in g\}\,,$$
we have
$$\tilde{g}=\hat{V}^{-1}g^{-1}\hat{V}=\hat{V}^{-1}g\hat{V}\,.$$
Eq~\eqref{dUVg1} and Eq.~\eqref{dUVg2} are one-to-one correspondent. If we have a method by $n$ gates to convert $\hat{V}$ to $\hat{U}$ under the gates set $g$ by the right-invariant way, then we have also a corresponding method by $n$ gates to convert $\hat{V}^{-1}$ to $\hat{U}^{-1}$ under the gates set $\tilde{g}$ by the right-invariant way. The converse is also true. For the left-invariant way $(\alpha=l)$, we have a similar conclusion. Thus, we have the following equality
\begin{equation}\label{eqdUVinv}
  d_\alpha(g;\hat{U},\hat{V})=d_\alpha(\tilde{g};\hat{U}^{-1},\hat{V}^{-1})\,.
\end{equation}
Here $\tilde{g}=\hat{V}^{-1}g^{-1}\hat{V}$ for $\alpha=r$ and $\tilde{g}=\hat{V}g^{-1}\hat{V}^{-1}$ for $\alpha=l$.
This is a fundamental symmetry for relative complexity in quantum circuits.

Similar to the discussion in Sec.~\ref{sec42}, for discrete qubit systems, we have
$$\tilde{g}\neq g$$
in general  so
$$d_\alpha(g;\hat{U},\hat{V})\neq d_\alpha(g;\hat{U}^{-1},\hat{V}^{-1})\,.$$
However, if we consider the SU($n$) group, as what we have argued in Sec.~\ref{sec42}, the gates set should be replaced by Lie algebra $\mathfrak{su}(n)$. Because of $\mathfrak{su}(n)=\hat{V}^{-1}\mathfrak{su}(n)\hat{V}=\hat{V}\mathfrak{su}(n)\hat{V}^{-1}$, a symmetry Eq.~\eqref{eqdUVinv} suggests that we should have
\begin{equation}\label{eqdUVinv2}
  d_\alpha(\mathfrak{su}(n);\hat{U},\hat{V})=d_\alpha(\mathfrak{su}(n);\hat{U}^{-1},\hat{V}^{-1})\,.
\end{equation}
%
In geometrical theory of complexity, the relative complexity in a SU($n$) group is naturally defined by the minimal length of curves connecting two operators such as
\begin{equation}\label{disUVg1}
  d_\alpha(\hat{U},\hat{V}):=\min\{L_\alpha[c]|\forall c, s.t., c(0)=\hat{V},~c(1)=\hat{U}\}\,,
\end{equation}
where we omit the argument $\mathfrak{su}(n)$ in the relative complexity.
Eq.~\eqref{eqdUVinv2} shows that the inverse map $\x\mapsto\x^{-1}$ should be a ``distance-preserving map'', i.e.,
\begin{equation}\label{disUVg2}
  d_\alpha(\hat{U},\hat{V})=d_\alpha(\hat{U}^{-1},\hat{V}^{-1})\,.
\end{equation}
The Finsler version of the \textit{Myers-Steenrod theorem} says that every surjective distance-preserving map in a Finsler manifold is also an isometric map (see the theorem 3.2 of chapter 3 in Ref.~\cite{deng2012homogeneous}). Thus, we see that the inverse map should be an isometry of complexity geometry. As a result, we have
\begin{equation}\label{inverLa}
  \forall c,~~~L_\alpha[c]=L_\alpha[c^{-1}]\,.
\end{equation}
This shows that the Myers-Steenrod theorem with basic symmetry~\eqref{eqdUVinv} can guarantee the path-reversal symmetry for SU($n$) groups. Then according to Eq.~\eqref{pathCPT1}, we find that $\F(H)=\F(\hat{U} H\hat{U}^{-1})$ and $\F(H)=\F(-H)$. This argument gives us the third method to show bi-invariance and the second method to show reversibility.

\subsection{Equivalence between ``bra-world'' and ``ket-world''}\label{othermetric2}
To describe a quantum system in a pure state, we usually use a ``ket'' state vector $|\cdot\rangle$. The time evolution of the system is governed by a unitary operator $c(t)$:
\begin{equation}\label{ketct1}
  |\psi(t)\rangle=c(t)|\psi(0)\rangle\,.
\end{equation}
However, it is artificial to choose the ``ket'' vector to present physics and we can equivalently use the ``bra''  state vector $\langle\cdot|$. In the ``bra-world'', the time evolution is given by
\begin{equation}\label{ketct2}
  \langle\psi(t)|=\langle\psi_0|c(t)^{\dagger}=\langle\psi_0|c(t)^{-1}\,.
\end{equation}
%


If $c(t)$, the curve in SU($n$) group, presents the time evolution of a system in ``ket-world'', then $c(t)^{-1}$ presents the time evolution of the same system in ``bra-world''.
Because all the physics should be invariant under the change of our formalism from ``ket-world'' to ``bra-world'' it is natural to expect that the ``length(cost)'' of $c(t)$ and $c(t)^{-1}$ are also the same:
\begin{equation} \label{uio}
L_\alpha[c]=L_\alpha[c^{-1}]\,.
\end{equation}
Starting with Eq. \eqref{uio}, we can derive other symmetries of complexity. This offers the fourth method to show bi-invariance and the third method to show reversibility.

\section{{Invariance of the cost function}}\label{lastapp}

It seems that the complexity is invariant under some transformation is weaker than the requirement that the cost function is invariant under that transformation. In this appendix we explain indeed they are equivalent. 

The argument is based on \textit{Myers-Steenrod theorem}, which says that every surjective distance-preserving map in a Finsler manifold is also a diffeomorphism (the theorem 3.2 of chapter 3 in Ref.~\cite{deng2012homogeneous}). 

Therefore, by  \textit{Myers-Steenrod theorem},  if we can show that the transformations relevant to \eqref{adjointcs0}, \eqref{arugueA1s1}, and \eqref{path-inver} are surjective distance-preserving maps, then we can say that they are diffeomorphisms so the cost is also invariant because those transformations are just diffeomorphisms.

Now we only need to prove that the transformations relevant to \eqref{adjointcs0}, \eqref{arugueA1s1}, and \eqref{path-inver} are surjective {\it distance}-preserving maps. 
For the proof, let us consider arbitrary two points $\V$ and $\W$ in an SU($n$) group manifold and 
denote the {\it distance} (shortest geodesic length) from $\V$ to $\W$ by $d(\V,\W)$.  

Note that the Finsler geometry is required to be right-invariant or left-invariant, which makes a connection between the distance and the complexity as follows 
\begin{equation}
d_r(\V,\W)=d_r(\I,\W\V^{-1})=\C_r(\W\V^{-1})\,,
\end{equation}
where the subscript $r$ means we consider the right-invariant case. We explain only for the right invariant case because a left-invariant case works similarly.

For \eqref{adjointcs0}, we only need to show that the unitary transformation preserve the distance i.e.
\begin{equation}
d_r(\V,\W)=d_r(\U\V\U^{-1},\U\W\U^{-1}) \,.
\end{equation}
It can be shown as follows. 
\begin{equation} \label{uT1}
\begin{split}
d_r(\V,\W)&=\C_r(\W\V^{-1})=\C_r(\U\W\V^{-1}\U^{-1}) \\
&=\C_r(\U\W\U^{-1}(\U\V\U^{-1})^{-1})=d_r(\U\V\U^{-1},\U\W\U^{-1})\,,
\end{split}
\end{equation}
where for the second equality we used the fact that the complexity is invariant under a unitary transformation. 
By \textit{Myers-Steenrod theorem} the unitary transformation is a diffeomorphism so $L[c(s)]=L[\U c(s)\U^{-1}]$.

For \eqref{path-inver} we only need to show that the distance is invariant under the inverse map $: \x\mapsto\x^{-1}$ i.e.
\begin{equation}
d_r(\V,\W)= d_r(\V^{-1},\W^{-1})) \,.
\end{equation}
It can be shown as follows. 
\begin{equation}
\begin{split}
d_r(\V,\W)&= \C_r(\W\V^{-1})=\C_r(\V\W^{-1})=\C_r(\W^{-1}\W\V\W^{-1})\\&=d_r(\W\V^{-1}\W^{-1},\W^{-1})=d_r(\V^{-1},\W^{-1})) \,,
\end{split}
\end{equation}
where for the second equality we used the fact that the complexity is invariant under an inverse map and for the fifth equality we used the fact that the distance is invariant under a unitary transformation with $\hat{U}=\W^{-1}$ in \eqref{uT1}.

Finally  \eqref{arugueA1s1} is also valid because of \eqref{path-inver} and \eqref{pathCPT1}.

\bibliographystyle{JHEP}

\begin{thebibliography}{10}

\bibitem{Ryu:2006bv}
S.~Ryu and T.~Takayanagi, \emph{{Holographic derivation of entanglement entropy
  from AdS/CFT}},
  \href{http://dx.doi.org/10.1103/PhysRevLett.96.181602}{\emph{Phys. Rev.
  Lett.} {\bf 96} (2006) 181602},
  [\href{http://arxiv.org/abs/hep-th/0603001}{{\tt hep-th/0603001}}].

\bibitem{Susskind:2014moa}
L.~Susskind, \emph{{Entanglement is not enough}},
  \href{http://dx.doi.org/10.1002/prop.201500095}{\emph{Fortsch. Phys.} {\bf
  64} (2016) 49--71}, [\href{http://arxiv.org/abs/1411.0690}{{\tt 1411.0690}}].

\bibitem{Susskind:2014rva}
L.~Susskind, \emph{{Computational Complexity and Black Hole Horizons}},
  \href{http://dx.doi.org/10.1002/prop.201500093,
  10.1002/prop.201500092}{\emph{Fortsch. Phys.} {\bf 64} (2016) 44--48},
  [\href{http://arxiv.org/abs/1403.5695}{{\tt 1403.5695}}].

\bibitem{Stanford:2014jda}
D.~Stanford and L.~Susskind, \emph{{Complexity and Shock Wave Geometries}},
  \href{http://dx.doi.org/10.1103/PhysRevD.90.126007}{\emph{Phys. Rev.} {\bf
  D90} (2014) 126007}, [\href{http://arxiv.org/abs/1406.2678}{{\tt
  1406.2678}}].

\bibitem{Brown:2015bva}
A.~R. Brown, D.~A. Roberts, L.~Susskind, B.~Swingle and Y.~Zhao,
  \emph{{Holographic Complexity Equals Bulk Action?}},
  \href{http://dx.doi.org/10.1103/PhysRevLett.116.191301}{\emph{Phys. Rev.
  Lett.} {\bf 116} (2016) 191301}, [\href{http://arxiv.org/abs/1509.07876}{{\tt
  1509.07876}}].

\bibitem{Cai:2016xho}
R.-G. Cai, S.-M. Ruan, S.-J. Wang, R.-Q. Yang and R.-H. Peng, \emph{{Action
  growth for AdS black holes}},
  \href{http://dx.doi.org/10.1007/JHEP09(2016)161}{\emph{JHEP} {\bf 09} (2016)
  161}, [\href{http://arxiv.org/abs/1606.08307}{{\tt 1606.08307}}].

\bibitem{Lehner:2016vdi}
L.~Lehner, R.~C. Myers, E.~Poisson and R.~D. Sorkin, \emph{{Gravitational
  action with null boundaries}},
  \href{http://dx.doi.org/10.1103/PhysRevD.94.084046}{\emph{Phys. Rev.} {\bf
  D94} (2016) 084046}, [\href{http://arxiv.org/abs/1609.00207}{{\tt
  1609.00207}}].

\bibitem{Chapman:2016hwi}
S.~Chapman, H.~Marrochio and R.~C. Myers, \emph{{Complexity of Formation in
  Holography}}, \href{http://dx.doi.org/10.1007/JHEP01(2017)062}{\emph{JHEP}
  {\bf 01} (2017) 062}, [\href{http://arxiv.org/abs/1610.08063}{{\tt
  1610.08063}}].

\bibitem{Carmi:2016wjl}
D.~Carmi, R.~C. Myers and P.~Rath, \emph{{Comments on Holographic Complexity}},
  \href{http://dx.doi.org/10.1007/JHEP03(2017)118}{\emph{JHEP} {\bf 03} (2017)
  118}, [\href{http://arxiv.org/abs/1612.00433}{{\tt 1612.00433}}].

\bibitem{Reynolds:2016rvl}
A.~Reynolds and S.~F. Ross, \emph{{Divergences in Holographic Complexity}},
  \href{http://dx.doi.org/10.1088/1361-6382/aa6925}{\emph{Class. Quant. Grav.}
  {\bf 34} (2017) 105004}, [\href{http://arxiv.org/abs/1612.05439}{{\tt
  1612.05439}}].

\bibitem{Kim:2017lrw}
R.-Q. Yang, C.~Niu and K.-Y. Kim, \emph{{Surface Counterterms and Regularized
  Holographic Complexity}},
  \href{http://dx.doi.org/10.1007/JHEP09(2017)042}{\emph{JHEP} {\bf 09} (2017)
  042}, [\href{http://arxiv.org/abs/1701.03706}{{\tt 1701.03706}}].

\bibitem{Carmi:2017jqz}
D.~Carmi, S.~Chapman, H.~Marrochio, R.~C. Myers and S.~Sugishita, \emph{{On the
  Time Dependence of Holographic Complexity}},
  \href{http://dx.doi.org/10.1007/JHEP11(2017)188}{\emph{JHEP} {\bf 11} (2017)
  188}, [\href{http://arxiv.org/abs/1709.10184}{{\tt 1709.10184}}].

\bibitem{Kim:2017qrq}
R.-Q. Yang, C.~Niu, C.-Y. Zhang and K.-Y. Kim, \emph{{Comparison of holographic
  and field theoretic complexities for time dependent thermofield double
  states}}, \href{http://dx.doi.org/10.1007/JHEP02(2018)082}{\emph{JHEP} {\bf
  02} (2018) 082}, [\href{http://arxiv.org/abs/1710.00600}{{\tt 1710.00600}}].

\bibitem{Swingle:2017zcd}
B.~Swingle and Y.~Wang, \emph{{Holographic Complexity of
  Einstein-Maxwell-Dilaton Gravity}},
  \href{http://arxiv.org/abs/1712.09826}{{\tt 1712.09826}}.

\bibitem{Alishahiha:2015rta}
M.~Alishahiha, \emph{{Holographic Complexity}},
  \href{http://dx.doi.org/10.1103/PhysRevD.92.126009}{\emph{Phys. Rev.} {\bf
  D92} (2015) 126009}, [\href{http://arxiv.org/abs/1509.06614}{{\tt
  1509.06614}}].

\bibitem{Ben-Ami:2016qex}
O.~Ben-Ami and D.~Carmi, \emph{{On Volumes of Subregions in Holography and
  Complexity}}, \href{http://dx.doi.org/10.1007/JHEP11(2016)129}{\emph{JHEP}
  {\bf 11} (2016) 129}, [\href{http://arxiv.org/abs/1609.02514}{{\tt
  1609.02514}}].

\bibitem{Couch:2016exn}
J.~Couch, W.~Fischler and P.~H. Nguyen, \emph{{Noether charge, black hole
  volume, and complexity}},
  \href{http://dx.doi.org/10.1007/JHEP03(2017)119}{\emph{JHEP} {\bf 03} (2017)
  119}, [\href{http://arxiv.org/abs/1610.02038}{{\tt 1610.02038}}].

\bibitem{Aaronson:2016vto}
S.~Aaronson, \emph{{The Complexity of Quantum States and Transformations: From
  Quantum Money to Black Holes}},  2016.
\newblock \href{http://arxiv.org/abs/1607.05256}{{\tt 1607.05256}}.

\bibitem{WIP}
R.-Q. Yang, Y.-S. An, C.~Niu, C.-Y. Zhang and K.-Y. Kim, \emph{work in
  progress}, .

\bibitem{Nielsen1133}
M.~A. Nielsen, M.~R. Dowling, M.~Gu and A.~C. Doherty, \emph{Quantum
  computation as geometry},
  \href{http://dx.doi.org/10.1126/science.1121541}{\emph{Science} {\bf 311}
  (2006) 1133--1135},
  [\href{http://arxiv.org/abs/http://science.sciencemag.org/content/311/5764/1133.full.pdf}{{\tt
  http://science.sciencemag.org/content/311/5764/1133.full.pdf}}].

\bibitem{Nielsen:2006:GAQ:2011686.2011688}
M.~A. Nielsen, \emph{A geometric approach to quantum circuit lower bounds},
  {\emph{Quantum Info. Comput.} {\bf 6} (May, 2006) 213--262}.

\bibitem{Dowling:2008:GQC:2016985.2016986}
M.~R. Dowling and M.~A. Nielsen, \emph{The geometry of quantum computation},
  {\emph{Quantum Info. Comput.} {\bf 8} (Nov., 2008) 861--899}.

\bibitem{038798948X}
D.~Bao, S.-S. Chern and Z.~Shen, \emph{An Introduction to Riemann-Finsler
  Geometry}.
\newblock Springer New York, New York, 2000,
  \href{http://dx.doi.org/10.1007/978-1-4612-1268-3}{10.1007/978-1-4612-1268-3}.

\bibitem{9810245319}
Z.~Shen, \emph{Lectures on finsler geometry (Series on Multivariate Analysis)}.
\newblock Wspc, 2001.

\bibitem{xiaohuan2006an}
M.~Xiaohuan, \emph{An introduction to Finsler geometry}.
\newblock World Scientific, Singapore Hackensack, NJ, 2006.

\bibitem{asanov1985finsler}
G.~S. Asanov, \emph{Finsler Geometry, Relativity and Gauge Theories}.
\newblock Springer Netherlands, Dordrecht, 1985.

\bibitem{Brown:2017jil}
A.~R. Brown and L.~Susskind, \emph{{The Second Law of Quantum Complexity}},
  \href{http://arxiv.org/abs/1701.01107}{{\tt 1701.01107}}.

\bibitem{Jefferson:2017sdb}
R.~A. Jefferson and R.~C. Myers, \emph{{Circuit complexity in quantum field
  theory}}, \href{http://dx.doi.org/10.1007/JHEP10(2017)107}{\emph{JHEP} {\bf
  10} (2017) 107}, [\href{http://arxiv.org/abs/1707.08570}{{\tt 1707.08570}}].

\bibitem{Yang:2017nfn}
R.-Q. Yang, \emph{{A Complexity for Quantum Field Theory States and Application
  in Thermofield Double States}},  \href{http://arxiv.org/abs/1709.00921}{{\tt
  1709.00921}}.

\bibitem{Khan:2018rzm}
R.~Khan, C.~Krishnan and S.~Sharma, \emph{{Circuit Complexity in Fermionic
  Field Theory}},  \href{http://arxiv.org/abs/1801.07620}{{\tt 1801.07620}}.

\bibitem{nielsen2010quantum}
M.~Nielsen, \emph{Quantum computation and quantum information}.
\newblock Cambridge University Press, Cambridge New York, 2010.

\bibitem{Vanchurin:2017qii}
V.~Vanchurin, \emph{{Covariant Information Theory and Emergent Gravity}},
  \href{http://arxiv.org/abs/1707.05004}{{\tt 1707.05004}}.

\bibitem{Yang:2018tpo}
R.-Q. Yang, Y.-S. An, C.~Niu, C.-Y. Zhang and K.-Y. Kim, \emph{{More on
  complexity of operators in quantum field theory}},
  \href{http://arxiv.org/abs/1809.06678}{{\tt 1809.06678}}.

\bibitem{Latifi2013}
D.~Latifi and M.~Toomanian, \emph{On the existence of bi-invariant finsler
  metrics on lie groups},
  \href{http://dx.doi.org/10.1186/2251-7456-7-37}{\emph{Mathematical Sciences}
  {\bf 7} (Aug, 2013) 37}.

\bibitem{Latifi2011}
D.~Latifi and A.~Razavi, \emph{Bi-invariant finsler metrics on lie groups},
  {\emph{Journal of Basic and Applied Sciences} {\bf 5} (Nov., 2011) 607--511}.

\bibitem{higham2008functions}
N.~Higham, \emph{Functions of matrices : theory and computation}.
\newblock Society for Industrial and Applied Mathematics, Philadelphia, 2008.

\bibitem{PhysRevLett.115.180405}
G.~Evenbly and G.~Vidal, \emph{Tensor network renormalization},
  \href{http://dx.doi.org/10.1103/PhysRevLett.115.180405}{\emph{Phys. Rev.
  Lett.} {\bf 115} (Oct, 2015) 180405}.

\bibitem{PhysRevLett.115.200401}
G.~Evenbly and G.~Vidal, \emph{Tensor network renormalization yields the
  multiscale entanglement renormalization ansatz},
  \href{http://dx.doi.org/10.1103/PhysRevLett.115.200401}{\emph{Phys. Rev.
  Lett.} {\bf 115} (Nov, 2015) 200401}.

\bibitem{Miyaji:2015fia}
M.~Miyaji, T.~Numasawa, N.~Shiba, T.~Takayanagi and K.~Watanabe,
  \emph{{Continuous Multiscale Entanglement Renormalization Ansatz as
  Holographic Surface-State Correspondence}},
  \href{http://dx.doi.org/10.1103/PhysRevLett.115.171602}{\emph{Phys. Rev.
  Lett.} {\bf 115} (2015) 171602}, [\href{http://arxiv.org/abs/1506.01353}{{\tt
  1506.01353}}].

\bibitem{Molina-Vilaplana:2018sfn}
J.~Molina-Vilaplana and A.~del Campo, \emph{{Complexity Functionals and
  Complexity Growth Limits in Continuous MERA Circuits}},
  \href{http://arxiv.org/abs/1803.02356}{{\tt 1803.02356}}.

\bibitem{Chapman:2017rqy}
S.~Chapman, M.~P. Heller, H.~Marrochio and F.~Pastawski, \emph{{Towards
  Complexity for Quantum Field Theory States}},
  \href{http://arxiv.org/abs/1707.08582}{{\tt 1707.08582}}.

\bibitem{Caputa:2017urj}
P.~Caputa, N.~Kundu, M.~Miyaji, T.~Takayanagi and K.~Watanabe, \emph{{Anti-de
  Sitter Space from Optimization of Path Integrals in Conformal Field
  Theories}},
  \href{http://dx.doi.org/10.1103/PhysRevLett.119.071602}{\emph{Phys. Rev.
  Lett.} {\bf 119} (2017) 071602}, [\href{http://arxiv.org/abs/1703.00456}{{\tt
  1703.00456}}].

\bibitem{Caputa:2017yrh}
P.~Caputa, N.~Kundu, M.~Miyaji, T.~Takayanagi and K.~Watanabe, \emph{{Liouville
  Action as Path-Integral Complexity: From Continuous Tensor Networks to
  AdS/CFT}}, \href{http://dx.doi.org/10.1007/JHEP11(2017)097}{\emph{JHEP} {\bf
  11} (2017) 097}, [\href{http://arxiv.org/abs/1706.07056}{{\tt 1706.07056}}].

\bibitem{PhysRev.59.195}
G.~Randers, \emph{On an asymmetrical metric in the four-space of general
  relativity}, \href{http://dx.doi.org/10.1103/PhysRev.59.195}{\emph{Phys.
  Rev.} {\bf 59} (Jan, 1941) 195--199}.

\bibitem{deng2012homogeneous}
S.~Deng, \emph{Homogeneous Finsler spaces}.
\newblock Springer-Verlag New York, New York, N.Y, 2012.

\end{thebibliography}

\providecommand{\href}[2]{#2}\begingroup\raggedright\endgroup

\end{document}